\documentstyle[preprint,tighten,aps,epsf]{revtex}
\def\expfac{e^{-q^2/6\alpha^2}}
\def\hf{\frac{1}{2}}
\def\thf{\frac{3}{2}}
\def\twotrd{\frac{2}{3}}

\def\bpf{\vec p_{f}}
\def\bpi{\vec p_{i}}
\def\bpX{\vec p_{X}}
\def\bp{\vec p}
\def\bq{\vec q}

\def\bP{\vec P}
\def\bSigma{\mbox{\large{$\vec{\Sigma}$}}}
\def\bJ{\vec J}
\def\bi{\mbox{$i$}}

\def\bigrho{\mbox{\large $\rho$\normalsize}}

\def\bsigma{\mbox{\large{$\vec \sigma$}}}

\def\bra{\langle }
\def\ket{\rangle }
\def\sumX{\!\!\!\!\!\!\sum_{X=N^*,\Delta^*}}

\def\eq{\begin{equation}}
\def\ee{\end{equation}}
\def\eqa{\begin{eqnarray}}
\def\eea{\end{eqnarray}}
\def\dEx{\,E(q)-E_X(q)}
\def\dMx{\,M-M_X}

\newcommand\ca[2]{\,\bra\chi^{#1}_{_{#2}}|
\bsigma_3|\chi^{\lambda}_{_{m_N}}\ket\,}
\newcommand\cb[1]{\,\bra\chi^{\lambda}_{_{#1}}|
\bsigma_3|\chi^{\lambda}_{_{m_N}}\ket\,}

\newcommand\trija[6]{(-1)^{\hf+{#6}} \sqrt{2{#5}\!+\!1} 
\left(\begin{array}{ccc}\!\!{#1} &{#3} &{#5}\\ 
                         \!\!{#2} & {#4} & -{#6}\end{array}\!\!\right) }
\newcommand\trijb[6]{(-1)^{-\hf+{#6}} \sqrt{2{#5}\!+\!1} 
\left(\begin{array}{ccc}\!\!{#1} &{#3} &{#5}\\ 
                         \!\!{#2} & {#4} & -{#6}\end{array}\!\!\right) }
\def\6j#1#2#3#4#5#6{\mbox{$\left\{\begin{array}{ccc}
#1 & #2 & #3 \\
#4 & #5 & #6
\end{array}
\right\}$}}
\def\cg#1#2#3#4#5#6{\mbox{$\langle #1~#2,#3~#4\vert #5~#6\rangle$}}

\begin{document}  
\preprint{MKPH-T-00-17, ECT$^*$-00-009} 
\draft
\tighten
\title{Generalized polarizabilities of the proton in a constituent
quark model revisited}
\author{B.\ Pasquini,$^1$
S.\ Scherer,$^2$ and D.\ Drechsel$^2$}
\address{
$^1$ ECT$^\ast$, European Centre for Theoretical Studies in Nuclear Physics 
and Related Areas, \\Strada delle Tabarelle 286, 
I-38050 Villazzano (Trento), Italy\\ 
$^2$Institut f\"ur Kernphysik, Johannes Gutenberg-Universit\"at,
J.\ J.\ Becher-Weg 45, D-55099 Mainz, Germany}
\date{August 22, 2000}
\maketitle
\begin{abstract}
   We study low-energy virtual Compton scattering off the proton
within the framework of a nonrelativistic constituent quark model.
   The Compton tensor is divided into two separately gauge-invariant
contributions.
   The first consists of the groundstate propagation in
the direct and crossed channels together with an appropriately
chosen term to satisfy gauge invariance.
   The residual part contains the relevant structure information
characterized by the so-called generalized polarizabilities.
   We discuss two different schemes to obtain the generalized 
polarizabilities from the residual term.
   Explicit predictions for the generalized polarizabilities are
presented for the Isgur-Karl model.
   Our results are compared with previous predictions in that model
as well as other approaches.   

\end{abstract}

\section{Introduction}
   The electromagnetic polarizabilities $\bar{\alpha}$ (electric) and
$\bar{\beta}$ (magnetic) of real Compton scattering (RCS) \cite{Klein_1955}
describe, to leading order in the photon frequency, the model-dependent 
response of a spin-0 or spin-1/2 system beyond the low-energy theorem (LET)
\cite{Thirring_1950,Low_1954,GellMann_1954}.
   Within a classical framework these quantities are accessible to an intuitive
interpretation as a ``measure of the stiffness or rigidity of a system
\cite{Holstein_1990}.''
   There have been considerable experimental efforts to determine the
proton polarizabilities from Compton scattering off the proton 
\cite{Federspiel_91,Zieger_92,Hallin_93,MacGibbon_95}.
  Until recently, the most precise values for the proton polarizabilities were
derived in the work of MacGibbon {\it et al.}~\cite{MacGibbon_95} and
analyzed 
by means of dispersion relations  at fixed $t$ \cite{Lvov}.
The results were
 $\bar{\alpha}_p=(12.1\pm0.8\pm 0.5)\times 10^{-4}$ fm$^3$
and $\bar{\beta}_p=(2.1\mp 0.8 \mp 0.5) \times 10^{-4}$ fm$^3.$
   The analysis of Ref.\ \cite{MacGibbon_95} made use of the 
Baldin sum rule~\cite{Baldin_1960} which relates the sum of the two 
polarizabilities to the total photoabsorption cross section.
 The sum rule constraint was
$\bar{\alpha}_p+\bar{\beta}_p=(14.2\pm 0.5)\times 10^{-4}$ fm$^3$
\cite{Damashek_70}, while more recent analyses found
$(13.69 \pm 0.14)\times 10^{-4}$ fm$^3$ \cite{Babusci_98a}
and $[14.0\pm(0.3-0.5)]\times 10^{-4}$ fm$^3$
\cite{Levchuk_2000}.
   New real-Compton-scattering data below pion-production threshold 
have been measured at the Mainz Microtron (MAMI).
   The new global fit including these data results in the values  
$\bar{\alpha}_p=(12.24\pm0.24\pm 0.54\pm 0.37)\times 10^{-4}$ fm$^3$
and 
$\bar{\beta}_p=(1.57 \pm0.25 \pm 0.52\pm 0.37)\times 10^{-4}$ fm$^3$
\cite{Olmos_2000}.
   As there is no free neutron target the experimental information on the
neutron polarizabilities is much less certain.
   Results for the electric polarizability have been obtained from
low-energy neutron scattering off the Coulomb field of a heavy nucleus
\cite{Schmiedmayer_88,Koester_88,Schmiedmayer_91,Koester_95}.
   Alternatively, the quasifree Compton scattering reaction $d(\gamma,\gamma'
n)p$ \cite{Rose_90a,Rose_90b,Kolb_2000} as well as   
elastic deuteron Compton scattering \cite{Hornidge_2000}
have been investigated to extract information on the neutron polarizabilities. 

   Clearly, the concept of polarizabilities is open to generalizations in, at
least, two directions.
   One possibility consists of investigating higher-order terms 
in the expansion of the RCS amplitude. 
   For example, for a spin-1/2 system  one finds 
four additional constants (spin polarizabilities) at third order
\cite{Choudhury_1969,Lin_1971,Ragusa_1993,Babusci_98b}
and four terms for the spin-averaged 
amplitude at fourth order \cite{Babusci_98b,Fearing_98}.
   Another option is to allow at least one photon to be virtual.
   This second generalization, which was already discussed in the 
1970's in the context of nuclei \cite{Arenhoevel_1974}, has  
recently been applied to the nucleon \cite{GLT,GV}.   
 
   As in all studies with electromagnetic probes, the inclusion of virtual 
photons substantially increases the possibilities to 
investigate the structure of the target.
   The use of a virtual photon allows one to access longitudinal degrees of 
freedom and to vary the three-momentum and energy
transfer to the target independently.
   The potential of investigating electron-proton bremsstrahlung as a
source of the virtual Compton effect amplitude was already noticed
by Berg and Lindner \cite{Berg_1961} in the 1960's.
   In its most general form for one real and one virtual photon,
this amplitude can be described by twelve form factors of three invariants
\cite{Berg_1961}.
   A generalized low-energy theorem (GLET) analogous to the LET of RCS
\cite{Thirring_1950,Low_1954,GellMann_1954} was derived in Ref.\
\cite{Scherer_96}, where it was shown that, up to and including
second order in the momenta, all twelve amplitudes for the proton
can be expressed in terms of the proton mass
$M$, the anomalous magnetic moment $\kappa$, the electromagnetic Sachs
form factors $G_E$ and $G_M$, the mean square electric radius $r^2_E$, 
and the RCS electromagnetic polarizabilities
$\bar{\alpha}$  and $\bar{\beta}$. 

   In  Ref.\ \cite{GLT}, the model-dependent response beyond the
LET was analyzed by means of a multipole expansion.
   Only terms contributing to first order in the frequency of the
outgoing real photon were kept, and the result was expressed in
terms of ten generalized polarizabilities (GPs) which are functions of
the three-momentum of the virtual photon in the initial state.
   Further progress has been made with respect to implementing the constraints
due to the discrete symmetries in combination with particle crossing.
   In Refs.\ \cite{Drechsel_1997,Drechsel_1998b,Drechsel_1998c} it was shown
that only six of the originally ten GPs are independent
if charge-conjugation symmetry is combined with particle crossing.

   Following the very first calculation in the framework of a 
nonrelativistic quark model \cite{GLT,LTG}, there have
been numerous predictions for the GPs within various approaches,
including phenomenological Lagrangians 
\cite{Vanderhaeghen_1996,Korchin_1998},
the linear sigma model \cite{Metz_96}, 
chiral perturbation theory \cite{Hemmert_97}, 
the Skyrme model \cite{Kim_97},
a relativistic constituent quark model
\cite{Pasquini_1998}, 
and the so-called small scale expansion \cite{Hemmert_2000}.
   On the experimental side, first evidence for virtual Compton scattering
events was reported in Ref.\ \cite{Brand_1995}.
   The first results for two structure functions involving linear
combinations of GPs at $Q^2=0.33$ GeV$^2$
have been obtained from a dedicated VCS experiment
at MAMI \cite{Roche_2000}.
   Further experiments probing the GPs at different
values of momentum transfer are underway at Jefferson Lab \cite{Audit_1993}
and MIT-Bates \cite{Shaw_1997}.

   Our work is organized as follows. 
   We start out in Sec.\ II with a general discussion of the (virtual) Compton 
scattering tensor for a nonrelativistic system of $N$ particles.
   We propose a separation into two individually gauge-invariant pieces.
   The first part consists of the groundstate propagating in the direct and
crossed channels, supplemented by an appropriately chosen term to satisfy
gauge invariance.
   The residual part contains the structure information contributing to
the GPs.
   Section III deals with the multipole expansion of the Compton tensor
and the definition of the GPs according to Ref.\ \cite{GLT}.
   Two schemes are presented to identify those pieces
of the residual term that actually yield contributions to the GPs.
   In Sec.\ IV we reconsider the calculation of the GPs in the framework
of the Isgur-Karl model, 
   and Sec.\ V contains a short summary and some conclusions.      
   The more technical details can be found in the appendices.

\section{Hadronic Compton scattering tensor}

In this section we discuss the formalism to describe both real and virtual
Compton scattering $\gamma^\ast(\omega,\vec{q}\,)+N(E_i,\vec{p}_i)\to
\gamma^\ast(\omega',\vec{q}\,')+N(E_f,\vec{p}_f)$
off a composite system, denoted by $N$, within the framework of 
nonrelativistic quantum mechanics.
We only consider the main steps to derive the hadronic tensor of Compton 
scattering and refer the interested reader to 
Refs.~\cite{Petrunkin_64,Ericson_73,Friar_74,Friar_75,Arenhoevel_85} 
and the more recent work of Refs.~\cite{Lvov_93,Scherer_99} 
for a detailed discussion.

   The starting point is the nonrelativistic Hamiltonian for a composite 
system of $N$ particles with masses $m_\alpha$,
\eqa
\label{eq:h0}
H_0=-\sum_{\alpha=1}^N
\frac{\left(\stackrel{\rightarrow}\nabla_\alpha\right)^2}
{2m_\alpha}+\sum_{\alpha<\beta} V_{\alpha\beta},
\eea
where $\vec r_\alpha$ and $-i\stackrel{\rightarrow}\nabla_\alpha$ refer,
respectively, to
the position and conjugate momentum of the particle $\alpha$ in the coordinate
representation.
   We use natural units, i.e., $\hbar=c=1$, $e>0$, and $\alpha=e^2/4\pi
\approx 1/137$.
   For simplicity, 
we only consider a local potential $V_{\alpha\beta}$, i.e., the interaction 
between the constituents does not contain momentum-dependent and/or exchange 
forces, and thus avoid the problem of exchange currents.  
   The eigenstates of Eq.\ (\ref{eq:h0}), denoted by $|X\vec{p}_X\rangle$,
are normalized according to
\begin{equation}
\label{eq:normalization}
 \bra X' \vec p_{X'}|X \vec p_X\ket=\delta^3 (\vec p_{X'}-\vec p_{X})
\delta_{X'X},
\end{equation}
   where $X$ and $\vec{p}_X$ refer to internal quantum numbers and the
total momentum of the system, respectively.

   The interaction with an external electromagnetic field is introduced 
via minimal substitution.
   In addition, we include a coupling to an intrinsic magnetic
dipole moment $\vec{\mu}_\alpha=e_\alpha\vec{\sigma}_\alpha/2m_\alpha$
of the constituents, where $\vec{\sigma}_\alpha/2$ is the Pauli spin operator.
   The resulting electromagnetic interaction Hamiltonian in the
Schr\"{o}dinger representation reads\footnote{We adopt 
a covariant notation, even though a calculation
within the framework of nonrelativistic quantum mechanics does not provide a
covariant result.}
\eqa
H_I&=&H_{I,\,1}+H_{I,\,2},\\
H_{I,\,1}&=&\int d^3 x\,  J^\mu(\vec x) A_\mu(\vec x),\\
H_{I,\,2}&=&\frac{1}{2}
\int d^3x
\int d^3 x'
B^{\mu\nu}(\vec x,\vec x\,')A_\mu(\vec x)
A_\nu(\vec x\,'),
\eea
where $A^\mu(\vec x)$ is the second-quantized photon field and 
\eqa
\label{j}
& &\vec J(\vec x)=
\sum_{\alpha=1}^N
\frac{ e_\alpha}{2m_\alpha }\left[\delta^3(\vec x-\vec r_\alpha)
\left(\frac{\stackrel{\rightarrow}\nabla_\alpha}{i}-\vec \sigma_\alpha\times
\stackrel{\rightarrow}\nabla_\alpha\right)
-\left(\frac{\stackrel{\leftarrow}\nabla_\alpha}{i}
+\vec \sigma_\alpha\times\stackrel{\leftarrow}\nabla_\alpha\right)
\delta^3(\vec x-\vec r_\alpha)\right],\nonumber\\
& &\\
& &\rho(\vec x)=\sum_{\alpha=1}^{N}e_\alpha \delta^3(\vec x-\vec r_\alpha),\\
\label{rho}
& &
B^{\mu 0}=B^{0\nu}=0,\quad B^{ij}(\vec x,\vec x\,')
=\delta_{ij}\sum_{\alpha=1}^N
\frac{e_\alpha^2}{m_\alpha}\delta^3(\vec x-\vec r_\alpha)
\delta^3(\vec x\,'-\vec r_\alpha).
\eea
   The total Hamiltonian $H=H_0+H_{rad}+H_I$, where $H_{rad}$ refers to the 
Hamiltonian of the radiation field, is time independent.

The hadronic Compton tensor is obtained by calculating the 
contributions of $H_{I,2}$ and $H_{I,1}$ in first-order and
second-order perturbation theory, respectively,\footnote{For notational
convenience we keep the four-momenta $q$ and $q'$ as arguments,   
although their time components are related by energy conservation.}
\eq
\label{scat_amplitude1}
M^{\mu\nu}_{fi}(q', q,\vec{p}\,)=
S^{\mu\nu}_{fi}(q', q)
+T^{\mu\nu}_{fi}( q', q,\vec{p}\,),
\ee
   where $q$ and $q'$ refer to the four-momenta of the
initial and final photons.
   In Eq.\ (\ref{scat_amplitude1}) we have also kept the dependence on a 
third independent three-momentum, namely, the average
of the initial and final target momenta $\vec{p}=(\vec{p}_i
+\vec{p}_f)/2$.
   Usually this dependence is reexpressed in terms of the photon three-momenta
when specifying the reference frame through $\vec{p}=\{(\vec{q}-\vec{q}\,')/2,
\vec{0},-(\vec{q}+\vec{q}\,')/2\}$ 
for the laboratory, Breit, and center-of-mass 
frame, respectively.
   This issue is of importance in the context of photon crossing 
as will be shown later.
   
   Our normalization convention for the Compton 
tensor follows from the S-matrix element 
of RCS, $\gamma(q,\lambda) 
+N(p_i,\sigma)\to\gamma(q',\lambda')
+N(p_f,\sigma')$:
\begin{equation}
S_{fi}= \delta^3(\vec{p}_f-\vec{p}_i)\delta^3(\vec{q}-\vec{q}\,')
\delta_{\sigma'\sigma}\delta_{\lambda'\lambda}
- \frac{i}{8\pi^2\sqrt{\omega\omega'}}\delta^4(p_f+q'-p_i-q) 
\epsilon_\mu'^\ast(\lambda')\epsilon_\nu(\lambda)
M^{\mu\nu}_{fi}(q', q,\vec{p}\,).
\end{equation}

   Using the interaction representation with respect to 
$H_0+H_{rad}$,\footnote{Alternatively, 
we could have treated the electromagnetic
field as a time-dependent external field and used the interaction 
representation with respect to $H_0$. 
   Both approaches yield the same result.}
   the first-order contribution is a sum of the ``seagull'' terms,
\eq
\label{seagull}
S^{\mu\nu}_{fi}(q', q)
= (2\pi)^3       \bra f \vec p_f|B^{\mu\nu}(q', q)
         |i \vec p_i\ket,
\ee
where 
\eqa
\label{bmunu}
B^{\mu\nu}(q', q)=\frac{1}{2}\int d^3z
\left[e^{-i\vec q\,'\cdot \vec z}B^{\mu\nu}(\vec z,0)
+ e^{i\vec q\cdot \vec z}
B^{\nu\mu}(\vec z,0)\right].
\eea
   The latter equation follows from translational
invariance,
$$B^{\mu\nu}(\vec{x},\vec{x}\,')=e^{-i \vec{P}\cdot \vec{x}\,'}
B^{\mu\nu}(\vec{x}-\vec{x}\,',0)e^{+i \vec{P}\cdot \vec{x}\,'},
$$
   where $\vec{P}$ refers to the total momentum operator of
the composite system.

   The second-order result from the direct and crossed channels reads  
\eqa
\label{s_u_ch}
T^{\mu\nu}_{fi}(q', q,\vec{p}\,)&=&
(2\pi)^3\sum_{X}\int d^3 p_X
         \bra f \vec p_f |J^{\mu}(0)|X \vec p_X\ket 
         \frac{(2\pi)^3\delta^3(\bpX-\bpi-\bq\,)}{E_f(\bpf)+\omega'-E_X(\bpX)} 
         \bra X \vec p_X|J^{\nu}(0)|i \vec p_i \ket
\nonumber\\
& &
+(2\pi)^3
\sum_{X}\int d^3 p_X
         \bra f \vec p_f|J^{\nu}(0)|X\vec p_X\ket 
         \frac{(2\pi)^3
\delta^3(\bp_X-\bpi+\bq\,')}{E_{f}(\bpf)-\omega-E_X(\bpX)} 
         \bra X \vec p_X|J^{\mu}(0)|i\vec p_i \ket,\nonumber\\  
\eea
   with the energy of the intermediate state $|X\vec{p}_X\rangle$ given by 
$E_X(\vec{p}_X)=\vec{p}\,^2_X/2M +\Delta E_X$, $M$ being the target mass
and $\Delta E_X$ the excitation energy of the corresponding state.
   For further reference we note that Eqs.\ (\ref{seagull}) and 
(\ref{s_u_ch}) are symmetrical under photon crossing,
$q^\mu\leftrightarrow -q'^\mu$ and $\nu\leftrightarrow \mu$.\footnote{
Sometimes the symmetry property $B^{\mu\nu}(\vec{x},\vec{x}\,')=
B^{\nu\mu}(\vec{x}\,',\vec{x})$ is used to express Eq.\ (\ref{bmunu})
as $\int d^3z
e^{-i\vec q\,'\cdot \vec z}B^{\mu\nu}(\vec z,0)$ which, however,
is no longer manifestly symmetrical under photon crossing.}

By use of the standard procedure 
to separate center-of-mass and internal motions 
\cite{Friar_74,Arenhoevel_85},
the matrix element of the current operator may be written as 
\eq
\label{eq:cm_intrinsic}
\bra f\bpf |J_\mu(0)|i\bpi\ket\equiv\frac{1}{(2\pi)^3}\bra f|
J_\mu(\vec q,\vec p\,)|i\ket,
\ee
where
\eqa
\label{eq:intrinsic}
\vec J(\vec q,\vec p\,)&=&
\vec j^{\,in}(\vec q\,)+\frac{\vec p}{M}\rho(\vec q\,),\nonumber\\
J_0(\vec q,\vec p\,)&=&\rho(\vec{q}\,),
\eea
with
$\vec q=\bpf-\bpi$.
   In Eq.\ (\ref{eq:intrinsic}), the intrinsic current operator
$\vec j^{in}(\vec q\,)$ and the charge density operator
$\rho(\vec q\,)$ are 
\eqa 
\label{eq:intrinsic_curr}
\vec j^{in}(\vec q\,)&=&\sum_\alpha \frac{e_\alpha}{2m_\alpha}\left(
\{e^{i\vec q\cdot \vec r\,'_\alpha},\vec{p}\,'_\alpha\}+
i\vec \sigma_\alpha \times\vec q
e^{i\vec q\cdot \vec r\,'_\alpha}\right),\\
\label{eq:intrinsic_charge}
\rho(\vec q\,)&=&\sum_\alpha e_\alpha 
e^{i\vec q\cdot \vec r\,'_\alpha},
\eea
where $\{a,b\}\equiv ab+ba$ is the standard anticommutator, and 
$\vec r\,'_\alpha $
and 
$\vec{p}\,'_\alpha$ are the intrinsic coordinates and momenta of the 
particles relative to the center of mass,
\eqa
\vec r\,'_\alpha &=& \vec r_\alpha - \vec R,\quad
\vec{R}=\frac{1}{M}\sum_\alpha m_\alpha \vec{r}_\alpha,\\
\vec{p}\,'_\alpha &=& \vec{p}_\alpha -\frac{m_\alpha}{M}\vec{P},\quad
\vec{P}=\sum_\alpha \vec{p}_\alpha.
\eea
   We note that the intrinsic coordinates and momenta do not satisfy the
canonical commutation relations,
\begin{equation}
\label{eq:comrel}
[r\,'_{\alpha,i},p\,'_{\beta,j}]=
i\delta_{ij}\left(\delta_{\alpha\beta}-\frac{m_\beta}{M}\right).
\end{equation}

   From now on we assume that both initial and final states,
denoted by $|0\rangle$ and $\langle 0'|$, respectively,
correspond to the ground state of the system.
   However, we explicitly allow for a change in the spin projection.
   Inserting Eq.\ (\ref{eq:cm_intrinsic}) 
into Eqs.\ (\ref{seagull})--(\ref{s_u_ch}) and integrating over
the center-of-mass momentum, 
the Compton tensor of Eq.\ (\ref{scat_amplitude1}) can be cast into the 
form\footnote{In the following
we omit the subscript $fi$.}
\eqa
\label{eq:1}
T^{\mu\nu}(q',q,\vec{p}\,)& = &
         \sum_{X}
         \bra 0' |J^{\mu}(-\bq\,',2\bpf+\bq\,')|X \ket 
         \frac{1}{E_f(\bpf)+\omega'-E_X(\bpf+\bq\,')} 
         \bra X|J^{\nu}(\bq,2\bpi+\bq\,)|0 \ket
\nonumber\\
& &\nonumber\\
&&+
         \sum_{X}
         \bra 0'|J^{\nu}(\bq,2\bpf-\bq\,)|X\ket 
         \frac{1}{E_f(\bpf)-\omega-E_X(\bpf-\bq\,)} 
         \bra X|J^{\mu}(-\bq\,',2\bpi-\bq\,')|0 \ket,
\nonumber\\
& &\\
\label{eq:2}
S^{\mu0}
&=&S^{0\nu}=0,\quad
S^{ij}=
\delta_{ij}
         \bra 0' |
         \sum_{\alpha}\frac{e^2_\alpha}{m_\alpha} 
         \mbox{e}^{i(\vec q-\vec q\,')\cdot\vec r\,'_\alpha}
         |0 \ket.
\eea

   Before calculating the GPs,
we split the Compton tensor into two parts that are 
separately gauge invariant and symmetrical under photon crossing,
\eqa
M^{\mu\nu}=
\tilde M_P^{\mu\nu}+
\tilde M_R^{\mu\nu}.
\eea
The modified pole term $\tilde M_P^{\mu\nu}$ is defined as
\eqa
\tilde M_P^{\mu\nu}=T_P^{\mu\nu}+G^{\mu\nu},
\eea
   where the pole term $T_P^{\mu\nu}$ corresponds to the
contribution of the intermediate ground state 
in the direct and crossed channels and $G^{\mu\nu}$
reads 
\eqa
\label{eq:gmunu}
G^{\mu 0}(q',q)
= 
G^{0 \nu}(q',q)=0,
\quad
G^{ij}(q', q)=
\delta_{ij}\frac{1}{M}\bra 0'|\rho(-\vec q\,')|0'\ket
\bra 0|\rho(\vec q\,)|0\ket \delta_{M_f M_i}.
\eea
   A derivation of $G^{\mu\nu}$ is given in Appendix A. 
   In particular,  $\tilde M_P^{\mu\nu}$ generates the correct
Thomson amplitude in the limit $q,q'\to 0$.

The residual term $\tilde M_R^{\mu\nu}$
is then given by 
\eqa
\label{mrtmunu}
\tilde M_R^{\mu\nu}=T_R^{\mu\nu}+\tilde S^{\mu\nu},
\eea
where $T_R^{\mu\nu}$ is the contribution
of the excited states in the direct and crossed channels, while the modified
seagull term is $\tilde  S^{\mu\nu}=
S^{\mu\nu}-G^{\mu\nu}.$

\section{Residual Compton tensor and generalized polarizabilities}

   In this section we review the multipole expansion of
the residual tensor and define the GPs according to Ref.~\cite{GLT}.
   In this context, we generalize the discussion of Ref.~\cite{GLT}
by allowing for a general spin $I$ of the initial and final states, 
respectively.
   We then discuss two different schemes of evaluating the
residual term $\tilde M_R^{\mu\nu}$, the first one based on the
presentation of Ref.\ \cite{LTG} and the second one on a systematic 
$1/M$ expansion, which is capable of incorporating the 
constraints of photon-crossing symmetry to leading order in $M^{-1}$.

\subsection{Multipole expansion and generalized polarizabilities}
   The starting point of Ref.\ \cite{GLT} for the definition of the GPs 
is the multipole decomposition of the residual term,
which is to be evaluated in the center-of-mass frame defined by 
$\vec p_i+\vec q=\vec p_f+\vec q\,'=0$,\footnote{We repeat that $\omega$ and
$\omega'$ are related by energy conservation.}
\begin{eqnarray}
\label{mtilde_R_munu_multexp}
\lefteqn{\tilde{M}^{\mu\nu}_R(M_f,\omega',\vec{q}\,';M_i,\omega,\vec{q}\,)=}
\nonumber\\
&&
4\pi\sum_{\rho,L,M,\atop\rho',L',M'}
g_{\rho'\rho'} 
V^\mu(\rho' L' M',\hat {q}\,')
H^{(\rho'L'M',\rho L M)}_R(M_f,\omega',|\vec{q}\,'|;M_i,\omega,|\vec{q}\,|)
V^{\nu\ast}(\rho L M,\hat {q})g_{\rho\rho},
\end{eqnarray}
where $\{V^{\mu}(\rho L M,\hat {q})\}$ constitutes the four-dimensional basis
of the multipole expansion of four-vector fields as introduced 
in Appendix C of Ref.~\cite{GLT}.
   In particular, $\rho (\rho')$ denotes the type of the initial (final)
multipolarity ($\rho=0$ scalar, $\rho=1$ magnetic, $\rho=2$ electric,
and $\rho=3$ longitudinal) and $L,M$ $(L',M')$ refer to the quantum numbers of 
the total angular momentum of the initial (final) photon.
   Note that in Eq.\ (\ref{mtilde_R_munu_multexp}) the dependence on the
arguments $\vec{q}$ and $\vec{q}\,'$ also results from the dependence
of $\tilde{M}^{\mu\nu}_R$ of Eq.\ (\ref{mrtmunu}) on $\vec{p}$, given by
$\vec{p}=-(\vec{q}+\vec{q}\,')/2$ in the center-of-mass frame.
   This implies that a naive substitution $[\mu,\omega,\vec{q}\,]
\leftrightarrow[\nu,-\omega',-\vec{q}\,']$ in Eq.\ 
(\ref{mtilde_R_munu_multexp}) is no longer equivalent to photon 
crossing, which assumes that $\vec{p}$ is not affected by such a
transformation.
   This can also be seen from energy conservation, where ``naive''
photon crossing would imply
$$
\omega'=\omega +E(-\vec{q}\,)-E(-\vec{q}\,')
\mapsto
-\omega'+E(\vec{q}\,')-E(\vec{q}\,),
$$
which clearly contradicts the correct relation under crossing,
$\omega'\leftrightarrow -\omega$.
   We will come back to this point in subsection III.C.

   Using the orthogonality property of the basis vectors 
$V^{\mu}(\rho L M,\hat {q})$, the multipoles can be extracted from 
Eq.\ (\ref{mtilde_R_munu_multexp}) as
\eqa
\label{eq:multipoles}
\lefteqn{
H^{(\rho' L'M',\rho L M)}_R(M_f,\omega',|\vec{q}\,'|;M_i,\omega,|\vec{q}\,|)=}
\nonumber\\
&&\frac{1}{4\pi}\int\,{\rm d}\hat{q}~{\rm d}\hat{q}'~
V^{\ast}_{\mu}(\rho' L' M',\hat {q}')
\tilde M_R^{\mu\nu}(M_f,\omega',\vec{q}\,';M_i,\omega,\vec{q}\,)
V_{\nu}(\rho L M,\hat{q}).
\eea
   Since the residual term is gauge invariant,  
$q'_\mu \tilde{M}^{\mu\nu}_R=0=q_\nu \tilde{M}^{\mu\nu}_R$,
it is sufficient to consider scalar, magnetic, and electric
multipoles only \cite{GLT},
\begin{eqnarray}
\label{mtilde_R_munu_multexp2}
\lefteqn{\tilde{M}^{\mu\nu}_R(M_f,\omega',\vec{q}\,';M_i,\omega,\vec{q}\,)=}
\nonumber\\
&&
4\pi\sum_{\rho,\rho'=0}^2\sum_{L,M,\atop L',M'}
g_{\rho'\rho'} 
W^\mu(\rho' L' M',\hat {q}\,')
H^{(\rho'L'M',\rho L M)}_R(M_f,\omega',|\vec{q}\,'|;M_i,\omega,|\vec{q}\,|)
W^{\nu\ast}(\rho L M,\hat {q})g_{\rho\rho},
\end{eqnarray}
where 
\begin{eqnarray*}
W^{\mu}(\rho L M,\hat {q})&=&
V^{\mu}(\rho L M,\hat {q})
+\delta_{\rho 0}\frac{\omega}{|\vec{q}\,|} V^{\mu}(3 L M,\hat {q}),
\quad \rho=0,1,2.
\end{eqnarray*}
   Finally, the dependence on the target spin projections is 
extracted by defining reduced multipoles,\footnote{The right-hand side of
Eq.~(65) of Ref.~\cite{GLT} contains a typographic error and should include
a summation over the projections $s$ \cite{Guichon_2000}.} 
\eqa
\label{eq:red_multipoles}
H^{(\rho' L',\rho L)S}_R(\omega',|\vec{q}\,'|;\omega,|\vec{q}\,|)&=&
\sum_{M_{i},\,M_{f}}
\sum_{M,\,M'}(-1)^{L+M+I+M_{f}}
\cg{I}{-M_{f}}{I}{M_{i}}
{S}{s}\nonumber\\
& &\times\cg{L'}{M'}{L}{-M}{S}{s}
H^{(\rho' L'M',\rho L M)}_R(M_{f},\omega',|\vec{q}\,'|;
M_i,\omega,|\vec{q}\,|),
\nonumber\\
\eea
where $I$ is the spin of the target, and any of the $2S+1$ projections $s$
can be chosen.
   The selection rules due to the conservation of total angular momentum and 
parity are
$$
\vert L-L'\vert\le S\le L+L',\quad
0\le S\le 2 I,\quad
(-1)^{\delta_{\rho 1}+L}=(-1)^{\delta_{\rho'1}+L'}.
$$

   From now on we assume the final-state photon to be real,
$\omega'=|\vec{q}\,'|$.
   The definition of the GPs requires to identify the 
leading-order behavior in $|\vec{q}\,'|$ for any given multipole.
   In the Siegert limit of $|\vec{q}\,'|\to 0$, the final-state
electric and longitudinal multipoles are related by
\begin{equation}
\label{siegert}
\lim_{|\vec{q}\,'|\to 0} H^{(2 L',\rho L)S}_R=
\lim_{|\vec{q}\,'|\to 0}
\sqrt{\frac{L'+1}{L'}} H^{(3 L',\rho L)S}_R,
\end{equation}
which in turn can be expressed in terms of  
$H^{(0 L',\rho L)S}_R$ by gauge invariance.
   The leading-order behavior of a scalar or magnetic multipole
of order $L'$ is given by $|\vec{q}\,'|^{L'}$.
   The treatment of the initial-state photon is similar, except 
that one is interested in the behavior of the multipole
for arbitrary values of $|\vec{q}\,|$.
   The general relation between electric and longitudinal multipoles is
given by 
\eqa
\label{ely}
\vec{\cal E}^L_M(\hat{q})=\vec{\cal L}^L_M(\hat{q})
+\sqrt{\frac{2L+1}{L}}\vec{{\cal Y}}^{L\,\,L+1}_M(\hat{q}),
\eea
   where $\vec{\cal E}^L_M(\hat{q})$, $\vec{\cal L}^L_M(\hat{q})$, and 
$\vec {\cal Y}^{JL}_{M}(\hat q)$
are electric and longitudinal vectors of the multipole expansion, 
and vector spherical harmonics, respectively.
   While the first term on the right-hand side of Eq.\ (\ref{ely}) can be
expressed by the scalar multipole, 
the second term leads to the so-called mixed multipoles,
which are neither of electric
nor longitudinal type \cite{GLT},
\eqa
\label{eq:el_multipoles}
\lefteqn{\hat{H}^{(\rho' L'M',L M)}_R
(M_f,\omega',|\vec{q}\,'|;M_i,\omega,|\vec{q}\,|)=}\nonumber\\
&&\frac{1}{4\pi}\int\,{\rm d}\hat{q}~{\rm d}\hat{q}'~
V^{\ast}_{\mu}(\rho' L' M',\hat {q}')
\tilde M_R^{\mu i}(M_f,\omega',\vec{q}\,';M_i,\omega,|\vec{q}\,|)
\left[\vec {\cal Y}^{LL+1}_{M}(\hat q)\right]^i,
\eea 
with reduced multipoles analogous to 
Eq.\ (\ref{eq:red_multipoles}).
   The GPs of Ref.\ \cite{GLT} are now defined as
\begin{eqnarray}
\label{eq:GP}
P^{(\rho' L',\rho L)S}(|\vec{q}\,|) & = &
 \biggl[ \frac{1}{\omega^{\prime L'} |\vec{q}\,|^{L}}
   H_{R}^{(\rho' L',\rho L)S}(\omega', |\vec{q}\,|) \biggr]_{\omega'=0}
 \quad (\rho,\rho' = 0,1) \,,
 \nonumber \\
\hat{P}^{(\rho' L',L)S}(|\vec{q}\,|) & = &
 \biggl[ \frac{1}{\omega^{\prime L'} |\vec{q}\,|^{L+1}}
   \hat{H}_{R}^{(\rho' L',L)S}(\omega', |\vec{q}\,|) \biggr]_{\omega'=0}
 \quad (\rho' = 0,1) .
\end{eqnarray}
   Up to this point we have only incorporated constraints due to 
rotational symmetry, gauge invariance, and parity conservation.
   As such, the above analysis is valid in both a nonrelativistic and 
a relativistic framework.
   However, additional restrictions apply, once a covariant, field-theoretical 
approach is chosen. 
   In particular, if one assumes symmetry under particle crossing 
in combination with charge-conjugation invariance, one obtains
a set of four independent linear equations, involving the ten generalized 
polarizabilities with $L'=1$ \cite{Drechsel_1997,Drechsel_1998b}.

\subsection{Expansion of Liu, Thomas, and Guichon}
 
   Let us first reconsider the calculation of the GPs according to 
Ref.~\cite{LTG}, where the hadronic tensor 
$\tilde M^{\mu\nu}_R$ is split into a leading and a recoil contribution.
   For that purpose one defines
the matrix elements of the charge density and current
operators of Eqs.\
(\ref{eq:cm_intrinsic})--(\ref{eq:intrinsic_charge}) 
between the ground 
state and the intermediate states $X\ne 0$,
\begin{eqnarray}
\label{overlap_int}
\bigrho_X(\bq\,)&=&\sum_\alpha e_\alpha
\langle X|e^{i\vec q\cdot \vec r\,'_\alpha}|0\rangle,\nonumber\\
\bP_X(\vec q\,)&=&\sum_\alpha \frac{e_\alpha}{2m_\alpha}\langle X|
\{e^{i\vec q\cdot \vec r\,'_\alpha},\vec{p}\,'_\alpha\}
|0\rangle,\nonumber\\
\bSigma_X(\vec q\,)&=&\sum_\alpha\frac{e_\alpha}{2m_\alpha}
\langle X|
\vec\sigma_\alpha 
e^{i\vec q\cdot \vec r\,'_\alpha}|0\rangle.
\end{eqnarray}
   With this convention the current matrix elements of the direct-channel 
(subscript $d$) of Eq.\ (\ref{eq:1}) read\footnote{
In the following we suppress the spin indices in our notation. 
}  
\begin{eqnarray}
\label{eq:direct_me}
J^0_{d,X0}(\bq\,)&\equiv&\langle X|\rho(\vec{q}\,)|0\rangle
=\bigrho_X(\bq\,),\nonumber\\
\vec{J}_{d,X0}(\bq\,)&\equiv&\bra X|\bJ(\bq,-\bq\,)|0\ket 
=\bP_X(\vec q\,)+i\bSigma_X(\vec q\,)\times \vec q
-\frac{\vec q}{2M}
\bigrho_X(\vec q\,),\nonumber\\
J^0_{d,0X}(\bq\,')&\equiv&
\langle 0|\rho(-\vec{q}\,')|X\rangle
=\bigrho_X^*(\bq\,'),\nonumber\\
\vec{J}_{d,0X}(\vec q\,')&\equiv&
\bra 0|\bJ(-\vec q\,', -\vec q\,')|X\ket
=\bP_X^*(\bq\,')
-i\bSigma_X^*(\vec q\,')\times \vec q\,'
-\frac{\bq\,'}{2M}\bigrho_X^*(\bq\,').
\end{eqnarray}
   In the crossed channel (subscript $c$) the intermediate states propagate
with momentum $\vec p_X=-\vec q-\vec q\,'$, 
resulting in matrix elements depending on both
$\vec q$ and $\vec q \,'$. 
   As in Ref.~\cite{LTG}, we write these matrix elements as
\eqa
\bra X|\bJ(-\bq\,',-2\vec q-\vec q\,')|0\ket &\equiv&
\vec{J}_{c,X0}(\bq\,')+\delta \vec J_{c,X0}(\bq,\bq\,'),\nonumber
\\
\bra 0|\bJ(\vec q, -2\vec q\,'-\vec q\,)|X\ket
&\equiv&
\vec{J}_{c,0X}(\bq\,)+\delta \vec J_{c,0X}(\bq,\bq\,'), 
\eea
where 
\eqa
\label{jcrossed}
J^0_{c,X0}(\bq\,')&=&\bigrho_X(-\bq\,'),\nonumber\\
\vec{J}_{c,X0}(\bq\,')&=&
\bP_X(-\bq\,')
-i\bSigma_X(-\vec q\,')\times\vec q\,'
-\frac{\bq\,'}{2M}\bigrho_X(-\bq\,'),\nonumber\\
J^0_{c,0X}(\bq\,)&=&\bigrho_X^*(-\bq\,),\nonumber\\
\vec{J}_{c,0X}(\bq\,)&=&
\bP_X^*(-\bq\,)
+i\bSigma_X^*(-\vec q\,)\times\vec q
-\frac{\bq}{2M}\bigrho_X^*(-\bq\,),\nonumber\\
\delta J^0_{c,X0}&=&\delta J^0_{c,0X}=0,\nonumber\\
\delta \vec J_{c,X0}(\bq,\bq\,')&=&
-\frac{\vec q}{M}\bigrho_X(-\vec q\,'),\nonumber\\
\delta \vec J_{c,0X}(\bq,\bq\,')&=&
-\frac{\vec q\,'}{M}\bigrho_X^*(-\vec q\,).
\eea
    In Ref.\ \cite{LTG}, the leading part of the hadronic tensor is  
obtained by neglecting the $\delta \vec{J}_c$ terms of Eqs.\ (\ref{jcrossed}) 
in the crossed-channel matrix elements and by only taking account of the 
leading terms in the expansion of the denominators in a power series in 
$\omega'$,\footnote{In a nonrelativistic framework consistent with 
Eq.\ (\ref{eq:h0}) one would have $E_i(\vec q\,)-E_X(\vec q\,)=-\Delta E_X$.
   However, in Ref.\ \cite{LTG} the relativistic expressions 
$E_i(\vec q\,)=\sqrt{M^2+\vec{q}\,^2}$ and
$E_X(\vec q\,)=\sqrt{(M+\Delta E_X)^2+\vec{q}\,^2}$ have 
been used.}
\begin{eqnarray}
\label{denominator}
\frac{1}{E_f(-\vec{q}\,')+\omega'-E_X(\vec{0})}
&=&\frac{-1}{\Delta E_X}\left[1+\frac{\omega'}{\Delta E_X}+
{\cal O}(\omega'^2)\right],\nonumber\\
\frac{1}{E_i(-\vec q\,)-E_X(-\vec q-\vec q\, ')-\omega'}
&=&
\frac{1}{E_i(\vec q\,)-E_X(\vec q\,)}
\left[1+\frac{1}{E_i(\vec q\,)-E_X(\vec q\,)}
\left(\omega'+\frac{\vec q\cdot\vec q\,'}{E_X(\vec q\,)}\right)\right]
\nonumber\\
&&+{\cal O}(\omega'^2),
\end{eqnarray}
   where $\Delta E_X$ is the excitation energy of the state $X$. 
   As a result one finds [see Eq.\ (20) of Ref.\ \cite{LTG}]
\eqa
\label{eq:leading}
 \tilde{M}^{\mu\nu}_{R-LEADING}(\bq\,',\bq\,) = 
-\sum_{X\ne 0} \left[
     \frac{ J^{\mu}_{d,0X}(\bq\,')\, J^{\nu}_{d,X0}(\bq\,) }
{\Delta E_X}\right]
-
\sum_{X\ne 0} \left[
     \frac{ J^{\nu}_{c,0X}(\bq\,)\, J^{\mu}_{c,X0}(\bq\,') }
{E_X(\vec{q}\,)-E_i(\vec{q}\,)}\right]
+\tilde S^{\mu\nu}.
\eea
   The remaining contribution from the crossed channel is included in the 
recoil part, which also contains additional terms of first order
in $\omega'$ from the expansion of the energy denominators
in Eqs.\ (\ref{denominator}). 
   Accordingly, we obtain for the recoil part 
\begin{eqnarray}
\label{eq:recoil}
\lefteqn{\tilde{M}^{\mu\nu}_{R-RECOIL}(\bq\,',\bq\,)=}\nonumber\\ 
&&-\sum_{X\neq 0}\frac{\omega'}{(\Delta E_X)^2}
J^\mu_{d,0X}(\vec{q}\,') J^\nu_{d,X0}(\vec{q}\,)\nonumber\\
&&+\sum_{X\neq 0}\frac{1}{[E_X(\vec q\,)-E_i(\vec q\,)]^2}\left[
\omega'+\frac{\vec{q}\cdot\vec{q}\,'}{E_X(\vec q\,)}\right]
J^\nu_{c,0X}(\vec{q}\,)
[J^\mu_{c,X0}(\vec{q}\,')+\delta J^\mu_{c,X0}(\vec{q},\vec{q}\,')]\nonumber\\
&&-\sum_{X\neq 0}\frac{1}{E_X(\vec q\,)-E_i(\vec q\,)}
\left[J^\nu_{c,0X}(\vec{q}\,)
\delta J^\mu_{c,X0}(\vec{q},\vec{q}\,')
+\delta J^\nu_{c,0X}(\vec{q},\vec{q}\,')J^\mu_{c,X0}(\vec{q}\,')
+\delta J^\nu_{c,0X}(\vec{q},\vec{q}\,')
\delta J^\mu_{c,X0}(\vec{q},\vec{q}\,')\right]\nonumber\\
&&+{\cal O}(\omega'^2).
\end{eqnarray}
   This result differs from Eq.\ (21) of Ref.\
\cite{LTG} by the presence of the first sum on the right-hand side,
the $\delta J$ term of the second sum, 
and the $\delta J \delta J$ of the third sum of
Eq.\ (\ref{eq:recoil}).
   However, this difference has no bearing on the 
calculation of the ten GPs with $L'=1$.
   In this case we only need to analyze terms
of Eq.\ (\ref{eq:recoil}) that are linear in $\omega'=|\vec{q}\,'|$ 
[see also Eqs.\ (\ref{eq:GP})]. 
   Furthermore, it follows from Eqs.\ (\ref{eq:multipoles}) and 
(\ref{eq:el_multipoles}) that the projection involves angular integrals
of the type $\int {\rm d} \hat{q}' Y^\ast_{1M'}(\hat{q}')\tilde{M}^{0\nu}_R$
and $\int {\rm d} \hat{q}' Y^\ast_{1m'}(\hat{q}')\tilde{M}^{i\nu}_R$
for scalar and magnetic final dipole radiation, respectively.
   The recoil contribution of Eq.\ (\ref{eq:recoil})
only modifies the GPs with a magnetic photon in the final state,  
as observed in Ref.\ \cite{LTG}.
   In fact, for $\mu=0$ we need to consider $J^0_{d,0X}$, $J^0_{c,X0}$, and
$\delta J^0_{c,X0}$.
   According to Eq.\ (\ref{jcrossed}), the last term vanishes identically.
   The first two terms have to be always evaluated at $\vec{q}\,'=0$,
because they are multiplied by expressions that 
are either manifestly or implicitly [$\delta \vec{J}_{c,0X}(\vec{q},
\vec{q}\,')$] of order $\omega'$.
   From Eqs.\ (\ref{eq:direct_me}) and (\ref{jcrossed}) we see that
the corresponding overlap integrals vanish due to the orthogonality of the 
excited states with respect to the ground state.
   In conclusion, the $L'=1$ scalar GPs do not
receive a contribution from Eq.\ (\ref{eq:recoil}).
   For the $L'=1$ magnetic GPs we need to consider $\mu\neq 0$. 
   In this case the terms proportional to $\omega'$ do not contribute, 
because the angular
integral of $Y^\ast_{1m'}$ multiplied by a $\vec{q}\,'$-independent
function vanishes.  
   Furthermore, since $\delta \vec{J}_{c,0X}(\vec{q},\vec{q}\,')$ is
of order $\omega'$, the last term 
in Eq.\ (\ref{eq:recoil}) containing $\delta \vec{J}_{c,X0}(\vec{q},\vec{0})$ 
vanishes due to orthogonality.
   In conclusion, one is left with
\begin{eqnarray}
\label{eq:recoil2}
\tilde{M}^{\mu\nu}_{R-RECOIL}(\bq\,',\bq\,)&=&
\sum_{X\neq 0}\frac{1}{[E_X(\vec q\,)-E_i(\vec q\,)]^2}
\frac{\vec{q}\cdot\vec{q}\,'}{E_X(\vec q\,)}
J^\nu_{c,0X}(\vec{q}\,) J^\mu_{c,X0}(\vec{q}\,')\nonumber\\
&&-\sum_{X\neq 0}\frac{1}{E_X(\vec q\,)-E_i(\vec q\,)}
\left[J^\nu_{c,0X}(\vec{q}\,)
\delta J^\mu_{c,X0}(\vec{q},\vec{q}\,')
+\delta J^\nu_{c,0X}(\vec{q},\vec{q}\,')J^\mu_{c,X0}(\vec{q}\,')\right]
\nonumber\\
&&+\cdots +{\cal O}(\omega'^2),
\end{eqnarray}
   where the ellipses denote terms which do not contribute to the 
$L'=1$ GPs.

\subsection{Nonrelativistic $1/M$ expansion}
   In this subsection we will consider a different procedure of ordering
the contributions of the excited states to the Compton tensor.
   Our aim is to analyze the implications of photon-crossing symmetry
for the GPs.
   We will see that some of the GPs will have to satisfy certain conditions
for $|\vec{q}\,|\to 0$.    
   At first, we will not specify the reference frame but 
both photons will be allowed to be virtual 
[see Eq.\ (\ref{eq:1}) for $X\neq 0$]. 
   Only at the end we will assume the initial and final photons
to be virtual and real, respectively, and restrict ourselves to the
center-of-mass frame.

   The starting point is a separation of four-current matrix elements
into an intrinsic current with respect to the center-of-mass system
and a center-of-mass convection current \cite{Arenhoevel_85}  [see Eqs.\ 
(\ref{eq:cm_intrinsic})--(\ref{eq:intrinsic_charge})]. 
   As an example, let us consider the following matrix element entering
the direct channel 
$$
\langle 0|J^{\mu}(-\vec{q}\,',2\vec{p}+\vec{q}\,)|X\rangle =
\langle 0|J^{\mu}(-\vec{q}\,',0)|X\rangle
+\langle 0|\delta J^{\mu}(-\vec{q}\,',2\vec{p}+\vec{q}\,)|X\rangle,
$$
with analogous expressions for the other combinations.
   Here,  we have defined for generic momenta $\vec{a}$ and $\vec{b}$
\begin{eqnarray*}
\mu=0:&& \langle 0|\delta J^{0}(\vec{a},\vec{b}\,)|X\rangle=0,\\
\mu\neq 0:&& \langle 0|\delta\vec{J}(\vec{a},\vec{b}\,)|X\rangle
=\frac{\vec{b}}{2M}\langle 0|\rho(\vec{a}\,)|X\rangle.
\end{eqnarray*}
   Using nonrelativistic kinematics $T(\vec{p}\,)=\vec{p}\,^2/(2M)$, we 
separate the energy denominators into excitation pieces
and kinetic contributions, 
\begin{eqnarray*}
\frac{1}{E_f(\vec{p}_f)+\omega'-E_X(\vec{p}_f+\vec{q}\,')}&=&
-\frac{1}{\Delta E_X}\left[1+\frac{\omega'}{\Delta E_X}
+\frac{T(\vec{p}_f)-T(\vec{p}_f+\vec{q}\,')}{\Delta E_X}
+\frac{\omega'^2}{(\Delta E_X)^2}+{\cal O}(3)\right],\\
\frac{1}{E_f(\vec{p}_f)-\omega-E_X(\vec{p}_f-\vec{q}\,)}&=&
-\frac{1}{\Delta E_X}\left[1-\frac{\omega}{\Delta E_X}
+\frac{T(\vec{p}_f)-T(\vec{p}_f-\vec{q}\,)}{\Delta E_X}
+\frac{\omega^2}{(\Delta E_X)^2}+{\cal O}(3)\right],
\end{eqnarray*}
where ${\cal O}(3)$ refers to terms which are suppressed by 
$1/(\Delta E_X)^3$,
$1/[(\Delta E_X)^2 M]$, etc.
   In this representation we have
\begin{eqnarray}
\label{eq:3}
\lefteqn{T^{\mu\nu}_{fi;X}(q',q,\vec{p}\,)=}\nonumber\\
&&         -\sum_{X\neq 0}
\frac{[\bra 0 |J^{\mu}(-\vec{q}\,',0)|X \ket
         + \bra 0 |\delta J^{\mu}(-\vec{q}\,',2\vec{p}+\vec{q}\,)|X \ket]
[\bra X|J^{\nu}(\vec{q},0)|0 \ket
         +\bra X|\delta J^{\nu}(\vec{q},2\vec{p}+\vec{q}\,')|0 \ket]}{
\Delta E_X}\nonumber\\
&&\quad\quad\quad \times
\left[1+\frac{\omega'}{\Delta E_X}
+\frac{T(\vec{p}_f)-T(\vec{p}_f+\vec{q}\,')}{\Delta E_X}
+\frac{\omega'^2}{(\Delta E_X)^2}+{\cal O}(3)\right]\nonumber\\
& &\nonumber\\
&&-\sum_{X\neq 0}
         \frac{[\bra 0|J^{\nu}(\vec{q},0)|X\ket 
+\bra 0|\delta J^{\nu}(\vec{q},2\vec{p}-\vec{q}\,')|X\ket]
[\bra X|J^{\mu}(-\vec{q}\,',0)|0 \ket
+\bra X|\delta J^{\mu}(-\vec{q}\,',2\vec{p}-\vec{q}\,)|0 \ket]}{
\Delta E_X}\nonumber\\
&&\quad\quad\times\left[1-\frac{\omega}{\Delta E_X}
+\frac{T(\vec{p}_f)-T(\vec{p}_f-\vec{q}\,)}{\Delta E_X}
+\frac{\omega^2}{(\Delta E_X)^2}+{\cal O}(3)\right].
\end{eqnarray}
   In order to expand Eq.\ (\ref{eq:3}) in powers of $1/M$, 
we introduce the following abbreviations for the direct-channel ($d$) terms:
\begin{eqnarray*}
A^\nu_d&=&\langle X|J^\nu(\vec{q},0)|0\rangle,\\
\delta A^\nu_d&=&\langle X|\delta J^\nu(\vec{q},2\vec{p}+\vec{q}\,')|0\rangle,
\\
B^\mu_d&=&\langle 0|J^\mu(-\vec{q}\,',0)|X\rangle,\\
\delta B^\mu_d&=&\langle 0|\delta J^\mu(-\vec{q}\,',2\vec{p}+\vec{q}\,)
|X\rangle,
\\
\Delta T_d&=&T(\vec{p}_f)-T(\vec{p}_f+\vec{q}\,').
\end{eqnarray*}
   The corresponding expressions for the the crossed channel ($c$) are obtained
by simply replacing $\vec{q}\leftrightarrow -\vec{q}\,'$ and $\mu
\leftrightarrow\nu$.
   We note that both $\delta$ and $\Delta$ terms count as order $1/M$. 

   The terms of leading order in $1/M$ and arbitrary order in
$1/\Delta E_X$,
\begin{equation}
\label{M0}
T^{\mu\nu}_{fi;X,l.o.}(q',q)=
-\sum_{X\neq 0} \left(\frac{B^\mu_d A^\nu_d}{\Delta E_X-\omega'}
+\frac{B^\nu_c A^\mu_c}{\Delta E_X+\omega}\right),
\end{equation}
   do not depend on $\vec{p}$.
   Equation (\ref{M0}) is symmetric under ``true photon crossing,'' 
but also under ``naive photon crossing'' in any frame, 
because it is independent of $\vec{p}$.
   As is shown in App.\ B, this property implies the following 
conditions on the leading behavior in $1/M$ for the reduced multipoles 
\begin{equation}
\label{crossinglom}
H^{(\rho' L',\rho L)S}_{R,l.o.}(\omega',|\vec{q}\,'|;\omega,|\vec{q}\,|)
=g_{\rho'\rho'}g_{\rho\rho}(-)^{L+L'-S} 
H^{(\rho L,\rho' L')S}_{R,l.o.}(-\omega,|\vec{q}\,|;-\omega',|\vec{q}\,'|).
\end{equation}

   Next we consider the $1/M$ corrections to Eq.\ (\ref{M0}) to arbitrary
order in $1/\Delta E_X$:
\begin{eqnarray}
\label{M-1}
&&-\sum_{X\neq 0} \left[
\frac{B^\mu_d \delta A^\nu_d+\delta B^\mu_d A^\nu_d}{\Delta E_X-\omega'}
+\frac{B^\nu_c \delta A^\mu_c+\delta B^\nu_c A^\mu_c}{\Delta E_X+\omega}
\right.\nonumber\\
&&\quad\quad\quad\left.+
\frac{B^\mu_d A^\nu_d}{\Delta E_X-\omega'}
\frac{\Delta T_d}{\Delta E_X-\omega'}
+\frac{B^\nu_c A^\mu_c}{\Delta E_X+\omega}
\frac{\Delta T_c}{\Delta E_X+\omega}
\right].
\end{eqnarray}
   The first line of Eq.\ (\ref{M-1}) contains the modifications of the
current matrix elements to first order in $1/M$ whereas the second line
involves the center-of-mass kinetic energies of the final and the
intermediate states, respectively.
   In this scheme, each order is separately photon-crossing symmetric.
   However, terms beyond leading order in $1/M$ will, in general, not be 
symmetric under naive photon crossing in the center-of-mass frame.

   We now turn to the case of real photons in the final state, 
$\omega'=|\vec{q}\,'|$, and explicitly specify the center-of-mass
frame.
   As we are not interested in terms beyond linear order in $\omega'$
and $1/M$, we make use of energy conservation, 
$$\omega\simeq \omega'-\frac{\vec{q}\,^2}{2M},
$$
   to expand the crossed-channel terms as
$$\frac{1}{\Delta E_X+\omega}=\frac{1}{\Delta E_X}\left[
1-\frac{\omega'}{\Delta E_X}+\frac{\vec{q}\,^2}{2M\Delta E_X}
-\frac{\omega'\vec{q}\,^2}{(\Delta E_X)^2 M}+{\cal O}(\omega'^2)\right].
$$
   We then obtain from Eq.\ (\ref{M0}) up to and including terms linear in 
$\omega'$ and $1/M$
\begin{eqnarray}
\label{M0exp}
T^{\mu\nu}_{fi;X,l.o.}(\vec{q}\,',\vec{q}\,)&=&
-\sum_{X\neq 0}\left[
\frac{1}{\Delta E_X}\left(1+\frac{\omega'}{\Delta E_X}\right)B^\mu_d A^\nu_d
+\frac{1}{\Delta E_X}
\left(1-\frac{\omega'}{\Delta E_X}\right)B^\nu_c A^\mu_c\right.
\nonumber\\
&&\left. +\frac{\vec{q}\,^2}{2M (\Delta E_X)^2} 
B^\nu_c A^\mu_c
-\frac{\omega'\vec{q}\,^2}{(\Delta E_X)^3 M} B^\nu_c A^\mu_c
\right]\nonumber\\
&&+{\cal O}(\omega'^2,1/M^2).
\end{eqnarray}
   Note that via energy conservation, Eq.\ (\ref{M0}) has also generated
terms in Eq.\ (\ref{M0exp}) which are of higher order in $1/M$.  

  For further evaluation of Eq.\ (\ref{M-1}) we use
$$\Delta T_d={ \cal O}(\omega'^2),\quad
\Delta T_c=-\frac{\vec{q}\,^2}{2M}-\frac{\vec{q}\cdot\vec{q}\,'}{M},
$$
and obtain
\begin{eqnarray}
\label{M-1exp}
T^{\mu\nu}_{fi;X,n.l.o.}(\vec{q}\,',\vec{q}\,)&=&
-\sum_{X\neq 0}\left[
\frac{1}{\Delta E_X}\left(1+\frac{\omega'}{\Delta E_X}\right)
(B^\mu_d\delta A^\nu_d+\delta B^\mu_d A^\nu_d)\right.\nonumber\\
&&\left.
+\frac{1}{\Delta E_X}\left(1-\frac{\omega'}{\Delta E_X}\right)
(B^\nu_c\delta A^\mu_c+\delta B^\nu_c A^\mu_c)\right.\nonumber\\
&&\left.
-\frac{\vec{q}\,^2}{2M} B^\nu_c A^\mu_c\frac{1}{(\Delta E_X)^2}
\left(1-2\frac{\omega'}{\Delta E_X}\right)
-\frac{\vec{q}\cdot\vec{q}\,'}{M}B^\nu_c A^\mu_c\frac{1}{(\Delta E_X)^2}
\right]\nonumber\\
&&+{\cal O}(\omega'^2,1/M^2).
\end{eqnarray}
   The last two terms of Eq.\ (\ref{M0exp}) cancel with two
corresponding terms in Eq.\ (\ref{M-1exp}) and the sum
can be written as
\begin{eqnarray}
\label{tensor_nr}
T^{\mu\nu}_{fi;X}(\vec{q}\,',\vec{q}\,)&=&
-\sum_{X\neq 0} \left[
\frac{B^\mu_d A^\nu_d}{\Delta E_X}
+\frac{B^\nu_c A^\mu_c}{\Delta E_X}
\right]
\nonumber\\
& &
-\sum_{X\neq 0} 
\frac{\omega'}{(\Delta E_X)^2}B^\mu_d A^\nu_d
\nonumber\\
& &+\sum_{X\neq 0} 
\frac{1}{(\Delta E_X)^2}\left[\omega'+\frac{\vec q\cdot \vec q\,'}{M}\right]
B^\nu_c A^\mu_c\nonumber\\
& &-\sum_{X\neq 0} \left[
\frac{B^\mu_d \delta A^\nu_d+\delta B^\mu_d A^\nu_d}{\Delta E_X}
+\frac{B^\nu_c \delta A^\mu_c+\delta B^\nu_c A^\mu_c}{\Delta E_X}
\right]\nonumber\\
& &\left.
-\sum_{X\neq 0} 
\frac{\omega'}{(\Delta E_X)^2}
\left(B^\mu_d \delta A^\nu_d+\delta B^\mu_d A^\nu_d
-B^\nu_c \delta A^\mu_c-\delta B^\nu_c A^\mu_c
\right)
\right]\nonumber\\
&&+{\cal O}(\omega'^2,1/M^2).
\end{eqnarray}
   As in the previous case we will now analyze which terms of Eq.\ 
(\ref{tensor_nr}) actually generate a contribution to the $L'=1$ GPs 
after angular integration with the spherical harmonics 
$Y^\ast_{1m'}(\hat{q}\,')$.
   Again, all terms explicitly proportional to $\omega'$ will not contribute,
because at leading order they are multiplied by expressions which do
not depend on the direction of $\hat{q}\,'$.
   The $\delta B^\mu_d$ term vanishes for $\mu=0$ and is of order $\omega'^2$
for $\mu\neq 0$, and one is left with
\begin{eqnarray}
\label{tensor_nrGP}
T^{\mu\nu}_{fi;X}(\vec{q}\,',\vec{q}\,)&=&
T^{\mu\nu}_{fi;X,LEADING}(\vec{q}\,',\vec{q}\,)
+T^{\mu\nu}_{fi;X,RECOIL}(\vec{q}\,',\vec{q}\,),
\end{eqnarray}
where
\begin{eqnarray}
\label{tensor_nrLEADING}
T^{\mu\nu}_{fi;X,LEADING}(\vec{q}\,',\vec{q}\,)&=&
-\sum_{X\neq 0} \left[
\frac{B^\mu_d A^\nu_d}{\Delta E_X}
+\frac{B^\nu_c A^\mu_c}{\Delta E_X}
\right],
\\
T^{\mu\nu}_{fi;X,RECOIL}(\vec{q}\,',\vec{q}\,)&=&
-\sum_{X\neq 0} \left[
\frac{B^\mu_d \delta A^\nu_d
+\delta B^\nu_c A^\mu_c}{\Delta E_X}\right]\nonumber\\
& &+\sum_{X\neq 0} 
\frac{1}{(\Delta E_X)^2}\frac{\vec q\cdot \vec q\,'}{M}
B^\nu_c A^\mu_c\nonumber\\
& &-\sum_{X\neq 0}
\frac{B^\nu_c \delta A^\mu_c}{\Delta E_X}\nonumber\\
&&+\cdots +{\cal O}(\omega'^2,1/M^2),
\label{tensor_nrRECOIL}
\end{eqnarray}   
   where the ellipses refer to terms which do not contribute to the 
$L'=1$ GPs.
   The terms in Eq.\ (\ref{tensor_nrLEADING}) generate
contributions to the GPs which result entirely from intrinsic currents.
   The first line of the recoil term in Eq.\ (\ref{tensor_nrRECOIL})
corresponds to $1/M$ corrections of the virtual-photon
absorption vertices in the direct and crossed channel, respectively.
   The second and third lines are $1/M$ corrections of the 
crossed-channel energy denominator and real-photon vertex, respectively.
   These last two corrections only affect GPs involving a magnetic
photon in the final state.

   Finally, by taking the nonrelativistic limit of the energies 
in the scheme of Ref.\ \cite{LTG},
\begin{eqnarray*}
E_i(\vec q\,)&=&\sqrt{M^2+\vec q\,^2}\simeq
M+\frac{\vec q\,^2}{2M},\\
E_X(\vec q)&=&=\sqrt{M_X^2+\vec q\,^2}
\simeq
M+\Delta E_X+\frac{\vec q\,^2}{2M},
\end{eqnarray*}
   it is straightforward to verify that the two expansion schemes
coincide up to and including terms of order $\omega'$ and 
$1/M$.
   They only differ in their separation into leading and recoil terms.

\section{Generalized polarizabilities in a NRCQM}
\label{NRCQM}

   In this section we discuss the GPs of the nucleon in the framework of a 
nonrelativistic system consisting of three constituent quarks.
   As in Ref.\ \cite{LTG}
we restrict ourselves to the
inclusion of the $\Delta(1232)$ resonance and the low-lying 
negative-parity baryons D$_{13}$(1520), S$_{11}$(1535), S$_{31}$(1620),
S$_{11}$(1650), S$_{13}$(1700), and D$_{33}$(1700).

\subsection{Matrix elements in the Isgur-Karl model}
   To be specific, we employ the model of Isgur and 
Karl~\cite{ISGURKARL} which describes the 
quark-quark potential by a harmonic-oscillator term plus a 
spin-dependent hyperfine interaction.
   The baryon states are expressed in a basis of SU(6) harmonic-oscillator 
wave functions, with the SU(6) multiplets generated by the combination 
of SU(2)$_{spin}$ and SU(3)$_{flavor}$ multiplets. 
   In particular, the nucleon and the $\Delta(1232)$ 
resonance belong to the ground-state spin-$1/2$ octet, $^{2}8$, and 
the spin-$3/2$ decuplet, $^{4}10$, of the $\underline{56}$ SU(6) 
supermultiplet, respectively.
   The multiplet of states associated with the negative-parity orbital 
excitation is classified in terms of a $\underline{70}$ supermultiplet of 
SU(6) which decomposes into $^{2}1$, $^{2}8$, $^{4}8$, and $^{2}10$ 
multiplets.
According to  Ref. \cite{ISGURKARL}, the strength of the hyperfine interaction
is fixed to reproduce the experimental mass
splitting of $N$ and $\Delta(1232)$ states,
while the remaining orbital excitations of the $\underline{70}$ multiplet
are constructed with mixing parameters 
describing the empirical spectrum quite well.

   Since we are interested in the non-strange sector only, we assume that
all three quarks have equal masses $m_q$.
   In addition, when calculating the matrix elements of the electromagnetic 
current of Eqs.\ (\ref{eq:intrinsic_curr}) and (\ref{eq:intrinsic_charge}), 
we take advantage of the overall 
symmetry of the SU(6) harmonic-oscillator wave function.
   This allows us to simplify the matrix elements of one-body operators,
\eqa
\label{eq:onebody}
\bra A |\sum_{i=1}^3 \hat O_i|B\ket=
3\bra A|\hat O_3|B\ket.
\eea
   As a result, the overlap integrals of 
Eq.\ (\ref{overlap_int}) can be written as
\begin{eqnarray}
\label{RHO}
 \rho_X(\vec q\,)&=&3\int d\vec{\rho} d\vec\lambda\,
   \mbox{e}^{-i\sqrt{\twotrd}\vec q\cdot \vec \lambda}\,
   \phi^{\dagger}_X\, \hat{Q}\, \phi_N,\\
\label{P}
\vec P_X(\vec q\,)&=&\sqrt{\frac{2}{3}}\frac{3}{2m_q}
\int d\vec\rho d\vec\lambda\,
   \mbox{e}^{-i\sqrt{\twotrd}\vec q\cdot\vec\lambda}\,
\phi^{\dagger}_X\,
   (i\stackrel{\rightarrow} \nabla_{\lambda}
  - i \stackrel{\leftarrow}\nabla_{\lambda} )
   \hat{Q}\,    \phi_N,\\
\label{SIGMA}
  \vec\Sigma_X(\vec q\,)&=&\frac{3}{2m_q} \int d\vec\rho d\vec\lambda\,
   \mbox{e}^{-i\sqrt{\twotrd}\vec q\cdot\vec\lambda}\,
   \phi^{\dagger}_X\,
   \vec\sigma_3 \hat{Q}\,
   \phi_N,
\end{eqnarray}
where 
\begin{eqnarray*}
\phi_N&=&
\phi_N(\vec \rho,\vec\lambda, {1\over 2}, M, 
        {1\over 2},\tau_{_N}),\\
\phi_X&=&\phi_X(\vec\rho,\vec\lambda, J_X,M_X,I_X,\tau_X)
\end{eqnarray*}
   denote the internal wave functions of the nucleon and
the excited states, respectively, with an obvious notation for
spin and isospin labels.
   We have introduced the standard Jacobi coordinates 
$
\vec \rho=(\vec r_1 - \vec r_2)/\sqrt 2$ and
$
\vec \lambda=(\vec r_1 + \vec r_2 - 2\vec r_3)/\sqrt 6,
$
and made use of
$\vec r\,'_3=\vec r_3 -\vec R=-\sqrt{2/3}\,\vec \lambda.$ 
   Furthermore, $\hat{Q} = (1/6+\tau_3/2)$ and $\vec\sigma_3$ 
denote the charge operator and the Pauli matrices of the third
quark, respectively.

   Explicit expressions for the contributions of the $P$-wave 
negative-parity states are given in Appendix \ref{app:1}.

\subsection{GPs in the framework of Liu, Thomas, and Guichon}

   We first discuss the results for the proton GPs that we obtain with the
same conventions used by Ref.~\cite{LTG} for the separation into leading and 
recoil terms.
   Following Ref.~\cite{ISGURKARL}, we use $m_q=350$ MeV for the quark mass 
and $\alpha=320$ MeV for the oscillator parameter.
   As was pointed out in Ref.~\cite{LTG}, the proton 
polarizabilities do not receive any contribution from 
the $N(^{4}8)$ multiplets.
   The mixing parameters $a_X$ encoding the $N(^{2}8)$ and $\Delta(^{2}10)$ 
composition of the resonant states are taken from Ref.~\cite{ISGURKARL} and 
are listed in Tab.\ I.

   With these assumptions we find for the leading contributions to the 
Compton tensor of Eq.\ (\ref{eq:leading}) 

\begin{eqnarray}
\label{p0101_s}
P^{(01,01)S} &=& {1 \over 18} {1 \over \alpha^2}\expfac  
       \sumX a_X^2   \left( {Z_d^{S,J_X} \over \dMx } 
      + {Z_c^{S,J_X} \over \dEx} \right), \hspace*{1.20cm} 
\\
\label{p0112_1}
P^{(01,12)1} &=& {1 \over 36} \sqrt{3 \over 5}  
   {1 \over m_q\alpha^2} \expfac  
     \sumX  a_X^2 {(-1)^{I_x-1/2} \over 2I_x}
       \left( {  Z_{ad}^{2,S,J_X} \over \dMx } +
     {  Z_{ac}^{2,S,J_X}   \over  \dEx} \right) ,
\\
\label{p1111_para}
P^{(11,11)S}_{para}&=&{4 \over 27}
{1 \over m_q^2}\expfac
  \left({ Z_\Delta^S\over M-M_\Delta }+(-1)^S
{ Z_\Delta^S\over E(q)-E_\Delta(q)} \right) ,
\hspace*{4.70cm}
\\
\label{p1111_dia}
P^{(11,11)S}_{dia} &=& \delta_{S0}{1 \over 3\sqrt{6}} 
      {1 \over m_q\alpha^2}\expfac .
 \hspace*{8.50cm}
\end{eqnarray}
   The angular coefficients $Z$ of the leading contributions 
are given in Tab.\ II.
   The diamagnetic term of Eq.\ (\ref{p1111_dia}) originates from the
modified seagull term $\tilde{S}^{\mu\nu}$ and contributes in the
spin-independent case only.
   The mixed GPs are given as the sum of two terms, corresponding to
the contributions from the convective ($ \hat{P}^{(01,1)S}_F$)
and spin ($\hat{P}^{(01,1)S}_\Sigma $) terms of the current at the 
virtual-photon vertex
\begin{eqnarray}
\hat{P}^{(01,1)S}&=& \hat{P}^{(01,1)S}_F + \hat{P}^{(01,1)S}_\Sigma, 
\\
\label{p011_s_f}
\hat{P}^{(01,1)S}_F &=&{\sqrt{2} \over 108\sqrt{3}}
    {1 \over m_q\alpha^2}\expfac \sumX a_X^2 
      \left( {Z_d^{S,J_X} \over \dMx } 
      + {Z_c^{S,J_X} \over \dEx} \right),
\\
\label{p011_sigma}
\hat{P}^{(01,1)S}_\Sigma &=&-{1 \over 36\sqrt{3}}
   {1 \over m_q\alpha^2} \expfac  
   \sumX a_X^2 {(-1)^{I_x-1/2} \over 2I_x}
    \left({ Z_{ad}^{1,S,J_X}  \over \dMx } 
       + {Z_{ac}^{1,S,J_X}  \over \dEx }\right).
\end{eqnarray}
   In the scheme of Ref.\ \cite{LTG} the spin-dependent
GPs $P^{(11,02)1}$, $P^{(11,00)1}$, and $\hat{P}^{(11,2)1}$, 
all of which lead to M1 radiation in the final state,  
vanish identically at leading order.

   The recoil corrections are exclusively generated by the crossed-channel
diagrams and only modify the GPs with a magnetic
final photon \cite{LTG} (see the discussion at the end of 
Sec.\ III.B). 
   To start with, the two GPs $P^{(11,00)1}$ and $P^{(11,02)1}$ receive a 
non-vanishing recoil contribution 
\begin{eqnarray}
\label{p1100_1}
P^{(11,00)1}_{recoil} &=& -  {1 \over \sqrt{3}\, }
      {q^2\over  m_q } \,\expfac
       \sumX {a_X^2 Z_{1100}^{J_X} \over E_X(q) [\dEx]^2 }  
       \left[1-  {E_X(q) [\dEx] \over 3\alpha^2 }\right],
\nonumber\\
& &\\
P^{(11,02)1}_{recoil} &=&   - {1 \over \sqrt{3}\,} {1\over m_q }\,\expfac
       \sumX {a_X^2 Z_{1102}^{J_X}\over E_X(q) [\dEx]^2 }  
       \left[1-  {E_X(q) [\dEx] \over 3\alpha^2 }\right] .
\nonumber\\
& &
\end{eqnarray}
   When discussing the recoil contribution to the remaining
three polarizabilities $ P^{(11,11)S}_{recoil}$ and 
$\hat{P}^{(11,2)1}_{recoil}$,
it is useful to distinguish between terms which result from
the spin-independent (C) and the spin-dependent ($\Sigma$) part of 
the virtual-photon vertex.
   We find 
\begin{eqnarray}
P^{(11,11)S}_{recoil}&=&P^{(11,11)S}_{recoil,C}+
P^{(11,11)S}_{recoil,\Sigma},
\\
 P^{(11,11)S}_{recoil,C}& =& 
 {1\over 18\,}
 {\alpha^2 \over  m_q^2 } \,\expfac
       \sumX { a_X^2 Z_{1111,C}^{S,J_X} \over E_X(q) [\dEx]^2 }
       \left[1-  {2E_X(q) [\dEx] \over 3\alpha^2 }\right] ,
    \hspace*{0.0cm}
\nonumber\\
& &\\
 P^{(11,11)S}_{recoil,\Sigma} &=& {1\over 18\,}{
q^2 \over  m_q^2 } 
\,\expfac
       \sumX a_X^2 {(-1)^{I_x-1/2} \over 2I_x}
     {Z_{1111,\Sigma}^{S,J_X} \over E_X(q) [\dEx]^2 }\nonumber\\
& &      \times \left[1-  {E_X(q) [\dEx] \over 3\alpha^2 }\right], 
    \hspace*{0.0cm}
\end{eqnarray}
 
\begin{eqnarray}
\label{p112_1_c}
\hat{P}^{(11,2)S}_{recoil}&=&\hat{P}^{(11,2)S}_{recoil,C}+
\hat{P}^{(11,2)S}_{recoil,\Sigma},
\\
\hat{P}^{(11,2)1}_{recoil,C} &=& 
 -{1 \over 6\sqrt{5}\,}{1 \over m_q^2 }\,\expfac
       \sumX {a_X^2  Z_{1102}^{J_X} \over E_X(q) [\dEx]^2 }  
       \left[1-  {E_X(q) [\dEx] \over 3\alpha^2 }\right] ,
\nonumber\\
& &\\
\label{p112_1_sigma}
\hat{P}^{(11,2)1}_{recoil,\Sigma} &=&  {1 \over 2\sqrt{5}\,}
{1 \over m_q^2 }\,\expfac
       \sumX 
a_X^2 {(-1)^{I_x-1/2} \over 2I_x}
{Z_{1102}^{J_X} \over E_X(q) [\dEx]^2 }\nonumber\\  
 & &   \label{p112_1}
\times   \left[1-  {E_X(q) [\dEx] \over 3\alpha^2 }\right] .
\end{eqnarray} 
   The values for the angular coefficients $Z$ of the recoil 
contributions are given in Tab.\ III.

\subsection{Comparison with Liu, Thomas, and Guichon}

   We now compare our results with those of 
Ref.~\cite{LTG}.\footnote{We stress that both calculations start from the
same model and use the same parameters and approximations.}
   As has been stated in Sec.\ III.A, the spin-dependent polarizabilities 
differ by an overall factor of three due to the different definition of the 
reduced multipoles in Eq.\ (\ref{eq:red_multipoles}).
   In contrast with Ref.~\cite{LTG}, we find, at leading order, 
two simple relations for the angular coefficients of the direct and crossed 
channels [see Eqs.\ (\ref{p0101_s})--(\ref{p011_sigma})], 
\begin{eqnarray}
\label{angular_relation1}
Z_c^{S,J_X}&=&(-1)^S Z_d^{S,J_X},\\
\label{angular_relation2}
Z_{ac}^{L,S,J_X}&=& (-1)^{S+1} Z_{ad}^{L,S,J_X}.
\end{eqnarray}
   In particular, as a result of Eq.\ (\ref{angular_relation1}) combined
with the relative phases between the direct- and crossed-channel 
contributions to $P^{(11,11)1}_{para}$,
the leading term of the $P^{(01,01)1}$ and  $P^{(11,11)1}$ polarizabilities
vanishes at the real-photon point. 
   This is in agreement with the constraint by photon crossing as 
derived in Eq.\ (\ref{Hredcross}) of App.\ B, but in contrast to the results
of Ref.\ \cite{LTG}.
   In addition, in our calculation the diamagnetic contribution 
$P^{(11,11)1}_{dia}$ is only 3/7 of the result of Ref.\ \cite{LTG}.
 We also find different expressions for $\hat P^{(01,1)S}$,
 in particular for the relative phase of the direct- and crossed-channel 
 contributions.
   Furthermore, the angular coefficients $Z^{J_X}_{1100}$
and $Z^{J_X}_{1102}$ occurring in $P^{(11,00)1}_{recoil}$ and
$P^{(11,02)1}_{recoil}$ are smaller by a factor of $1/2$.
   In addition, we find a recoil contribution to $P^{(11,11)S}_{recoil}$
for both spin-flip and no-spin-flip transitions, while
in Ref.~\cite{LTG} such a contribution is absent for $S=1$.

   The numerical results of Eqs.\ (\ref{p0101_s})--(\ref{p112_1})
are shown in Figs.\ 1 and 2 together with the calculation of 
Ref.~\cite{LTG}.
    For the generalized electric polarizability 
$$\alpha(|\vec{q}\,|)=-\frac{e^2}{4\pi}\sqrt{\frac{3}{2}} 
P^{(01,01)0}(|\vec{q}\,|),
$$
the two results are in agreement (full line of Fig.\ 1).
   The discrepancies in $P^{(01,01)1}$, $P^{(11,11)1}$, and  
$P^{(01,12)1}$ originate from the contributions of the crossed channel 
relative to the direct one.
   The different results for the generalized magnetic polarizability
$$\beta(|\vec{q}\,|)=-\frac{e^2}{4\pi}\sqrt{\frac{3}{8}} 
P^{(11,11)0}(|\vec{q}\,|)
$$
   are mainly due to the discrepancy in the calculation of the diamagnetic 
term, while the different evaluations of the recoil terms result only
in small deviations.
   Finally, the differences in calculating the 
         $\hat P^{(01,1)S}$ and $\hat P^{(11,2)1}_{recoil}$ 
polarizabilities give rise to discrepancies of almost one order of magnitude.
In particular, the $\hat P^{(11,2)1}_{recoil}$ polarizability receives its
main contribution from $\hat P^{(11,2)1}_{recoil,\Sigma}$ which 
has been neglected in Ref.~\cite{LTG}.

\subsection{Comparison with other calculations and experiment}
 
   In Figs.\ 3 and 4 we compare the two expansion schemes of 
Secs.\ III.B and III.C. 
   In each graph the solid line represents the full result according 
to the scheme of Liu, Thomas, and Guichon.
   Recall that in this framework the energy denominators of the
crossed channel contributions are written using relativistic
kinematics.
   The dashed line corresponds to a consistent nonrelativistic expansion up 
to and including terms linear in $1/M$. 
   The relevant expressions can be found in App.\ \ref{app2}.
   The contributions of leading order in $1/M$  
[see Eqs.\ (\ref{tensor_nrLEADING}) and (\ref{tensor_nrRECOIL})]
are separately displayed as dotted lines.
   
   First of all, we note that due to photon-crossing symmetry   
the leading contributions to both $P^{(01,01)1}$ and 
$P^{(11,11)1}$ vanish identically.
    The use of relativistic expressions in the energy denominators
for the crossed-channel terms leads to pronounced differences between
the two schemes, as soon as the leading term is vanishing or small.
   A striking example is given by the difference between the solid and dashed 
lines in $P^{(01,01)1}$ of Fig.\ 3,
which is entirely due to this different treatment of 
the crossed-channel energy denominators.
   Finally, at the real-photon point, the leading contributions of both
expansion schemes are equal, whereas the recoil terms differ by the 
contribution of second-order terms in $1/M$.

   Figures 5 and 6 display our results (full lines) together with the  
predictions of the linear sigma model \cite{Metz_96} (dashed lines),
an effective Lagrangian model \cite{Korchin_1998} (dotted lines),
and heavy-baryon chiral perturbation theory \cite{Hemmert_2000}
(dashed-dotted lines).

   An unpolarized measurement can be analyzed in terms of three structure
functions $P_{LL}$, $P_{TT}$, and $P_{LT}$ 
\cite{GLT,Drechsel_1998b,Roche_2000}
which are products of the GPs and the electromagnetic Sachs form 
factors $G_E$ and $G_M$,\footnote{The fourth structure function $P'_{LT}$ 
of Ref.\ \cite{GLT} 
is related to $P_{LT}$ if symmetry under particle crossing and
charge conjugation is applied \cite{Drechsel_1998b}.}
\begin{eqnarray}
\label{PLL} 
P_{LL}(|\vec{q}\,|) & = &
-2 \sqrt{6} M G_{E}(Q_{0}^{2}) P^{(01,01)0}(|\vec{q}\,|),
\\ 
\label{PTT}
P_{TT}(|\vec{q}\,|) & = &
\frac{3}{2} G_{M}(Q_{0}^{2})\left\{
2 \omega_{0} P^{(01,01)1}(|\vec{q}\,|)
 + \sqrt{2} |\vec{q}\,|^2 [
\sqrt{3} \hat{P}^{(01,1)1}(|\vec{q}\,|)
+P^{(01,12)1}(|\vec{q}\,|)]\right\}\nonumber\\
&=&3G_{M}(Q_{0}^{2})|\vec{q}\,|^2[\sqrt{2}P^{(01,12)1}(|\vec{q}\,|)
-P^{(11,11)1}(|\vec{q}\,|)/\omega_0],\\
\label{PLT}
P_{LT}(|\vec{q}\,|) & = &
\sqrt{\frac{3}{2}} \frac{M |\vec{q}\,|}{\sqrt{Q_{0}^{2}}}
G_{E}(Q_{0}^{2}) P^{(11,11)0}(|\vec{q}\,|)\nonumber\\
&&+\frac{\sqrt{3}}{2}\frac{\sqrt{Q_{0}^{2}}}{|\vec{q}\,|} G_{M}(Q_{0}^{2})
\left[ P^{(11,00)1}(|\vec{q}\,|)
+ \frac{|\vec{q}\,|^{2}}{\sqrt{2}} P^{(11,02)1}(|\vec{q}\,|) \right]\nonumber\\
&=&
\sqrt{\frac{3}{2}} \frac{M |\vec{q}\,|}{\sqrt{Q_{0}^{2}}}
G_{E}(Q_{0}^{2}) P^{(11,11)0}(|\vec{q}\,|)
+\frac{3}{2}\frac{\sqrt{Q_{0}^{2}}\,|\vec{q}\,|}{\omega_0} G_{M}(Q_{0}^{2})
P^{(01,01)1}(|\vec{q}\,|),
\end{eqnarray}
   where $\omega_0=\omega|_{\omega'=0}=M-\sqrt{M^2+|\vec{q}\,|^2}$ 
and $Q^2_0=-q^2|_{\omega'=0}=
-2M\omega_0$.
   We note that the second equations of Eqs.\ (\ref{PTT}) and(\ref{PLT}),
respectively, rely on symmetry under particle crossing and charge conjugation 
[see Eqs.\ (21c) and (21d) of Ref.\ \cite{Drechsel_1998b}]
which is not satisfied in the NRCQM.
   Table IV contains the predictions for the three response functions
for $|\vec{q}\,|=(0,240,600)$ MeV, corresponding to the real-photon point,
MIT-Bates \cite{Shaw_1997}, and MAMI \cite{Roche_2000} kinematics, 
respectively.
   For the Sachs form factors we used the parametrization of 
Ref.~\cite{Hoehler}.  
For $P_{TT}$ and $P_{LT}$ we quote both results obtained from 
Eqs.\ (\ref{PTT}) and (\ref{PLT}), respectively.
   
   Finally, in table V we compare the predictions with the first experimental
information obtained at MAMI \cite{Roche_2000} for the linear 
combination $P_{LL}-P_{TT}/\epsilon$ and $P_{LT}$  at $|\vec{q}\,|=600$ MeV.

\subsection{Particle crossing, charge conjugation, and gauge invariance} 
   The original definition of the GPs of Ref.\ \cite{GLT} was based on angular
momentum conservation, parity conservation, and gauge invariance.
   In Refs.\ \cite{Drechsel_1997,Drechsel_1998b} it was shown that only
six of the originally ten GPs are independent if
particle-crossing symmetry in combination with charge-conjugation invariance 
is imposed.
   As a result a set of four linear equations was obtained,
\begin{eqnarray}
\label{eq:rel1}
\hat{P}^{(01,1)0}&=&-\frac{\omega_0}{|\vec{q}\,|^2}
\left[\sqrt{\frac{2}{3}}P^{(01,01)0}+
\frac{1}{\sqrt{6}}P^{(11,11)0}\right],\\
& &\nonumber\\
\label{eq:rel2}
\hat{P}^{(11,2)1}&=&-\frac{1}{|\vec{q}\,|^2}\left[
\sqrt{\frac{2}{5}}P^{(11,11)1}+
\sqrt{\frac{3}{5}}\omega_0 P^{(11,02)1}\right],\\
& &\nonumber\\
\label{eq:rel3}
\hat{P}^{(01,1)1}&=
&-\frac{1}{|\vec{q}\,|^2}\left[\sqrt{\frac{2}{3}} \omega_0P^{(01,01)1}
+\sqrt{\frac{2}{3}}\frac{|\vec{q}\,|^2}{\omega_0} P^{(11,11)1}
-\frac{1}{\sqrt{3}}|\vec{q}\,|^2P^{(01,12)1}
\right],\\
& &\nonumber\\
\label{eq:rel4}
P^{(11,00)1}&=&\sqrt{3}\frac{|\vec{q}\,|^2}{\omega_0}P^{(01,01)1}
-\frac{1}{\sqrt{2}}
|\vec{q}\,|^2P^{(11,02)1},
\end{eqnarray}
   where $\omega_0=\omega|_{\omega'=0}$.  
   Following common use, we choose for the six independent GPs 
$\alpha$, $\beta$, $P^{(01,01)1}$, $P^{(11,11)1}$,
$P^{(01,12)1}$, and $P^{(11,02)1}$ as given in Figs.\ 1, 3, and 5.

   However, in a nonrelativistic framework particle crossing is not
a symmetry of the Compton tensor (see Sec.\ 2.3 of Ref.\ \cite{Scherer_99}).
   Thus one cannot expect the relations of Eqs.\ (\ref{eq:rel1}) - 
(\ref{eq:rel4}) to be satisfied in the NRCQM.
   Having defined $\Delta_i$ ($i=1,2,3,4$) as the difference between the 
left-hand and right-hand sides of Eqs.\ (\ref{eq:rel1}) - (\ref{eq:rel4})
normalized to their right-hand sides,
we show the discrepancies $\Delta_i$ in Fig.\ 7 as function
of $|\vec{q}\,|$.
   Clearly, the relations of Eqs.\ (\ref{eq:rel1}) - 
(\ref{eq:rel4}) are strongly violated on the average.

   Finally, another important limitation of the NRCQM is 
due to a violation of gauge invariance. 
   This problem can be traced back to essentially two causes: 
first, the Isgur-Karl model includes some effects of the anharmonic
terms in the  $qq$ potential only perturbatively in 
the energy but not in the wave functions. 
   Such a treatment leads to a mismatch between the resonance masses in the 
energy denominators of the Compton amplitude and the baryon states which 
enter in the current matrix elements.
   Second, the actual calculation truncates the configuration space
to only few intermediate states, while gauge invariance requires, in
principle, the full set of intermediate states.
   The well-known example in this context is, of course, the low-energy
Thomson limit which requires the inclusion of the sum over
all electric-dipole excitations (including the negative-energy states of
a relativistic theory).

\section{Summary and Conclusions}

   We discussed the general form of the virtual Compton scattering 
tensor for a nonrelativistic composite system.
   In particular, we focussed the attention on the generalized 
polarizabilities of the proton, defined from the multipole expansion of the 
Compton tensor.
   We performed a consistent nonrelativistic expansion of the 
structure-dependent amplitude allowing us to identify the constraints
due to photon crossing.
   As a model calculation, we reconsidered the proton GPs in a 
nonrelativistic constituent quark model.
   The model satisfies the constraint due to photon crossing at the
real photon point, but does not provide the relations among the
GPs due to nucleon crossing in combination with charge 
conjugation.
   As a consequence of its limitations regarding relativity, gauge 
invariance and chiral symmetry, the results of the model should be 
treated with some care. 
   There clearly is room for improvement in any of the above-mentioned
shortcomings.
   Nonetheless, the predictions provide an order-of-magnitude estimate
for the nucleon resonance contributions and as such 
are complementary to the results of the linear sigma
model and chiral perturbation theory emphasizing pionic
degrees of freedom and chiral symmetry.
   
\acknowledgements

   We would like to thank P.\ A.\ M.\ Guichon for useful comments
on the manuscript.
B.P. is grateful to A. Metz and M. Vanderhaeghen for stimulating discussions.
   This work was supported in part by the Deutsche Forschungsgemeinschaft
(SFB 443).
\appendix
\section{Gauge-invariant modified groundstate pole terms}
   Here we derive the result of Eq.\ (\ref{eq:gmunu}) for $G^{\mu\nu}$ 
which, in combination with the groundstate pole terms, constitutes the 
gauge-invariant tensor $\tilde{M}^{\mu\nu}_P$. 
   We start with the expression [see Eqs.\ (8a,b) of Ref.\ 
\cite{Friar_75} with the replacement $k\to q'$ 
and $k'\to -q$]  
$$
T^{\mu\nu}_{fi}(q',q,\vec{p}\,)=-i(2\pi)^3\int d^4 z e^{iq'\cdot z}
\langle 0 M_f\vec{p}_f|T[J^\mu(z) J^\nu(0)]|0 M_i \vec{p}_i\rangle,
$$
   where $J^\mu(x)=\exp(i H_0 t) J^\mu(\vec{x}) \exp(-i H_0 t)$.
   In general, we allow for a change in the spin projection from
$M_i$ to $M_f$.
   The groundstate pole contribution is given by
\begin{eqnarray}
\label{a:tmunup}
\lefteqn{T^{\mu\nu}_{P,fi}(q',q,\vec{p}\,)=}\nonumber\\
&&-i(2\pi)^3\sum_M\int d^3 P
\int d^4 z e^{iq'\cdot z}\left[
\theta(z_0)\langle 0 M_f\vec{p}_f|J^\mu(z)|0M\vec{P}\rangle
\langle 0M\vec{P}|J^\nu(0)|0M_i\vec{p}_i\rangle\right.\nonumber\\
&&\left.\hspace{12.5em}
+\theta(-z_0)\langle 0M_f\vec{p}_f|J^\nu(0)|0M\vec{P}\rangle
\langle 0M\vec{P}|J^\mu(z)|0 M_i\vec{p}_i\rangle\right].\nonumber\\
\end{eqnarray}
   The following procedure is very similar to the one used in deriving
Ward-Fradkin-Takahashi identities 
\cite{Ward_50,Fradkin_56,Takahashi_57} in Quantum Field Theory
[see, e.g., Chap. 6.1 of Ref.\ \cite{ChengLi}]. 
   Let us contract Eq.\ (\ref{a:tmunup}) with $q'_\mu$ (arguments suppressed),
$$q'_\mu T^{\mu\nu}_{P,fi}=
-i (2\pi)^3 \sum_M\int d^3 P
\int d^4 z \left(-i\partial_\mu   e^{iq'\cdot z}\right)\left[\cdots\right].
$$ 
   Symbolically this expression is of the type
\begin{displaymath}
\int d^4 z (\partial_\mu e^{iq'\cdot z})f^\mu(z)=
-\int d^4 z e^{iq'\cdot z}\partial_\mu f^\mu(z),
\end{displaymath}
where we made use of a partial integration, and assumed that the interaction 
is ``adiabatically'' switched on 
and off to get rid of the surface terms at $z_0=\pm\infty$.
   Similarly, use of the divergence theorem has been made.
   After applying this result to the above case,
$$q'_\mu T^{\mu\nu}_{P,fi}
= (2\pi)^3 \sum_M \int d^3 P \int d^4 z e^{iq'\cdot z}\partial_\mu
[\cdots ],
$$
we use the relation 
$$
\partial_\mu[ \theta(\pm z_0) g^\mu(z)]=\pm
\delta(z_0)g^0(z)+\theta(\pm z_0)\partial_\mu g^\mu(z),
$$
and obtain
\begin{eqnarray*}
q'_\mu T^{\mu\nu}_{P,fi}
&=&(2\pi)^3 \sum_M \int d^3 P \int d^4 z e^{iq'\cdot z}\left[
\theta(z_0)\langle 0M_f\vec{p}_f|\partial_\mu J^\mu(z)|0M\vec{P}\rangle
\langle 0M\vec{P}|J^\nu(0)|0M_i\vec{p}_i\rangle\right.\\
&&\hspace{11em}
+\theta(-z_0)\langle 0M_f\vec{p}_f|J^\nu(0)|0M\vec{P}\rangle
\langle 0M\vec{P}|\partial_\mu J^\mu(z)|0M_i\vec{p}_i\rangle\\
&&\hspace{11em}
+\delta(z_0)\langle 0M_f\vec{p}_f|\rho(z)|0M\vec{P}\rangle
\langle 0M\vec{P}|J^\nu(0)|0M_i\vec{p}_i\rangle\\
&&\hspace{11em}\left.
-\delta(z_0)\langle 0M_f\vec{p}_f|J^\nu(0)|0M\vec{P}\rangle
\langle 0M\vec{P}|\rho(z)|0M_i\vec{p}_i\rangle\right].
\end{eqnarray*}
   With $\partial_\mu J^\mu(z)=0$ as an operator identity, the first
two terms on the right-hand side of the equation vanish. 
   Performing the integration with respect to $z_0$, applying translational
invariance as $\rho(\vec{z})=\exp(-i\hat{\vec{P}}\cdot\vec{z})\rho(0)
\exp(i\hat{\vec{P}}\cdot\vec{z})$, integrating first with respect to 
$\vec{z}$ and then with respect to $\vec{P}$, the two remaining terms yield
\begin{eqnarray*}
q'_\mu T^{\mu\nu}_{P,fi}&=&(2\pi)^6\sum_M \left[
\langle 0M_f\vec{p}_f|\rho(0)|0M\vec{p}_f+\vec{q}\,'\rangle
\langle 0M\vec{p}_f+\vec{q}\,'|J^\nu(0)|0M_i\vec{p}_i\rangle\right.\\
&&\hspace{4em}\left.
-\langle 0M_f\vec{p}_f|J^\nu(0)|0M\vec{p}_i-\vec{q}\,'\rangle
\langle 0M\vec{p}_i-\vec{q}\,'|\rho(0)|0M_i\vec{p}_i\rangle\right].
\end{eqnarray*}
   Finally, the operator $\rho$ of Eq.\ (\ref{rho}) is diagonal in the spin 
projections and we obtain
\begin{eqnarray}
\label{qptmunup}
q'_\mu T^{\mu\nu}_{P,fi}
&=&(2\pi)^6\left[
\langle 0 M_f\vec{p}_f|\rho(0)|0M_f\vec{p}_f+\vec{q}\,'\rangle
\langle 0 M_f\vec{p}_f+\vec{q}\,'|J^\nu(0)|0M_i\vec{p}_i\rangle
\right.\nonumber\\
&&\hspace{3em}\left.
-\langle 0M_f\vec{p}_f|J^\nu(0)|0M_i\vec{p}_i-\vec{q}\,'\rangle
\langle 0M_i\vec{p}_i-\vec{q}\,'|\rho(0)|0M_i\vec{p}_i\rangle\right].
\end{eqnarray}
First we consider Eq.\ (\ref{qptmunup}) for $\nu=0$:
\begin{equation}
\label{nu=0}
q'_\mu T^{\mu0}_{P,fi}=
\left[\langle 0M_f|\rho(-\vec{q}\,')|0M_f\rangle
\langle 0M_f|\rho(\vec{q}\,)|0M_i\rangle
-\langle 0M_f|\rho(\vec{q}\,)|0M_i\rangle
\langle 0M_i|\rho(-\vec{q}\,')|0M_i\rangle\right]=0,
\end{equation}
where we inserted Eqs.\ (\ref{eq:cm_intrinsic}) and (\ref{eq:intrinsic_charge})
and made use of the diagonal nature of $\rho$.
   Similarly, inserting Eqs.\ (\ref{eq:cm_intrinsic}) -
(\ref{eq:intrinsic_charge}) for $\nu=j$ we obtain 
\begin{eqnarray}
\label{nu=j}
q'_\mu T^{\mu j}_{P,fi}
&=&\langle 0M_f|\rho(-\vec{q}\,')|0M_f\rangle\left[
\langle 0M_f|j^{in,j}(\vec{q}\,)|0M_i\rangle
+\frac{p_f^j+q\,'^j+p_i^j}{2M}
\langle 0M_f|\rho(\vec{q}\,)|0M_i\rangle\right]\nonumber\\
&&-\left[\langle 0M_f|j^{in,j}(\vec{q}\,)|0M_i\rangle
+\frac{p_f^j+p_i^j-{q\,'}^j}{2M}
\langle 0M_f|\rho(\vec{q}\,)|0M_i\rangle\right]
\langle 0M_i|\rho(-\vec{q}\,')|0M_i\rangle\nonumber\\
&=&\langle 0 M_f|j^{in,j}(\vec{q}\,)|0M_i\rangle\left[
\langle 0M_f|\rho(-\vec{q}\,')|0M_f\rangle
-\langle 0M_i|\rho(-\vec{q}\,')|0M_i\rangle
\right]\nonumber\\
&&+\frac{{q\,'}^j}{M}
\langle 0M_f|\rho(-\vec{q}\,')|0M_f\rangle
\langle 0M_f|\rho(\vec{q}\,)|0M_f\rangle\delta_{M_i M_f}\nonumber\\
&=&\frac{{q\,'}^j}{M}
\langle 0M_f|\rho(-\vec{q}\,')|0M_f\rangle
\langle 0M_i|\rho(\vec{q}\,)|0M_i\rangle\delta_{M_i M_f},
\end{eqnarray}
   where we made use of the fact that the groundstate matrix elements of 
$\rho$ are diagonal and do not depend on the projection.
   
   The calculation of $q_\nu T^{\mu\nu}_{P,fi}$ proceeds in a completely
analogous fashion. Equations (\ref{nu=0}) and (\ref{nu=j}) suggest to add
the term of Eq.\ (\ref{eq:gmunu}),
$$
G^{\mu 0}(q',q)
= 
G^{0 \nu}(q',q)=0,
\quad
G^{ij}(q', q)=
\delta_{ij}\frac{1}{M}\bra 0 M_f|\rho(-\vec q\,')|0 M_f\ket
\bra 0 M_i|\rho(\vec q\,)|0 M_i\ket \delta_{M_i M_f},
$$
with the result that $\tilde{M}^{\mu\nu}_P=T^{\mu\nu}_{P,fi}+
G^{\mu\nu}$ is gauge invariant. 
   In particular, $\tilde{M}^{\mu\nu}_P$ depends on groundstate
properties only.

\section{Photon crossing constraints at leading order in $1/M$}
   In Sec.\ III.C we have seen that the leading term of the
residual amplitude, in a $1/M$ expansion, satisfies the photon crossing
constraint [see Eq.\ (\ref{M0})]
\begin{equation}
\label{crossing}
T^{\mu\nu}_{fi;X,l.o.}(q',q)=
T^{\nu\mu}_{fi;X,l.o.}(-q,-q').
\end{equation}
   Because Eq.\ (\ref{crossing}) does not depend on the average target
momentum, it is also symmetric under ``naive photon crossing,''
$$
T^{\mu\nu}_{X,l.o.}
(M_f,\omega',\vec{q}\,';M_i,\omega,\vec{q}\,)=
T^{\nu\mu}_{X,l.o.}
(M_f,-\omega,-\vec{q};M_i,-\omega',-\vec{q}\,'),
$$
which in terms of the multipole expansion of Eq.\ (\ref{mtilde_R_munu_multexp})
implies 
\begin{eqnarray*}
&&
4\pi\sum_{\rho,L,M,\atop\rho',L',M'}
g_{\rho'\rho'} 
V^\mu(\rho' L' M',\hat {q}\,')
H^{(\rho'L'M',\rho L M)}_{X,l.o.}
(M_f,\omega',|\vec{q}\,'|;M_i,\omega,|\vec{q}\,|)
V^{\nu\ast}(\rho L M,\hat {q})g_{\rho\rho}\\
&=&4\pi\sum_{\rho,L,M,\atop\rho',L',M'}
g_{\rho\rho} 
V^\nu(\rho L M,-\hat {q})
H^{(\rho L M,\rho' L' M')}_{X,l.o.}
(M_f,-\omega,|\vec{q}\,|;M_i,-\omega',|\vec{q}\,'|)
V^{\mu\ast}(\rho' L' M',-\hat {q}\,')g_{\rho'\rho'}\\
&=&
4\pi\sum_{\rho,L,M,\atop\rho',L',M'}(-)^{L+M+L'+M'}
V^\mu(\rho' L' -M',\hat {q}\,')
H^{(\rho L M,\rho' L' M')}_{X,l.o.}
(M_f,-\omega,|\vec{q}\,|;M_i,-\omega',|\vec{q}\,'|)\\
&&\quad\quad\quad\times
V^{\nu\ast}(\rho L -M,\hat {q})g_{\rho'\rho'}.
\end{eqnarray*}
   In the last step we made use of 
$$ V^{\mu\ast}(\rho L M,-\hat{q})=g_{\rho\rho}(-)^{L+M}
V^\mu(\rho L -M,\hat{q}).
$$
   With the orthogonality property of the multipole basis 
$\{V^{\mu}(\rho L M,\hat {q})\}$ (see appendix C of Ref.~\cite{GLT})
we find
\begin{eqnarray}
\label{Hcross}
\lefteqn{H^{(\rho'L'M',\rho L M)}_{X,l.o.}
(M_f,\omega',|\vec{q}\,'|;M_i,\omega,|\vec{q}\,|)
=}\nonumber\\
&&(-)^{L-M+L'-M'}
g_{\rho'\rho'}g_{\rho\rho} 
H^{(\rho L -M,\rho' L' -M')}_{X,l.o.}
(M_f,-\omega,|\vec{q}\,|;M_i,-\omega',|\vec{q}\,'|).
\end{eqnarray}
   Because of the orthogonality relations of
Clebsch-Gordan coefficients we finally obtain for the reduced multipoles
\begin{eqnarray}
\label{Hredcross}
\lefteqn{H^{(\rho'L',\rho L)S}_{X,l.o.}
(M_f,\omega',|\vec{q}\,'|;M_i,\omega,|\vec{q}\,|)
=}\nonumber\\
&&(-)^{L+L'-S}
g_{\rho'\rho'}g_{\rho\rho} 
H^{(\rho L,\rho' L')S}_{X,l.o.}
(M_f,-\omega,|\vec{q}\,|;M_i,-\omega',|\vec{q}\,'|).
\end{eqnarray}
   We stress that the above derivation holds only for those pieces
that are independent of $\vec{p}$. 
   It is only in this case that true photon crossing is equivalent
to naive photon crossing in the center-of-mass frame.

\section{Overlap integrals of the current operator in the NRCQM}
\label{app:1}
The contribution of the SU(3) multiplets to the overlap integrals
defined in Eqs.\ (\ref{RHO})-(\ref{SIGMA}) is given by
($q=|\vec{q}\,|$)

\begin{eqnarray}
\label{RHO2}
 \bigrho_{_{N(^28)}}(\bq)&  = & - \bi  \frac{\sqrt{8\pi}}{3}  
    \frac{q}{\alpha} \expfac  \tau_{_N} 
    \trija{1}{M_X\!\!-\!m_{_N}}{\hf}{m_{_N}}{J_X}{M_X} 
    Y_{1M_X-m_{_N}}^*(\hat{q}),\nonumber\\
&&\\
\label{P2}
  \bP_{_{N(^28)}}(\bq)& = & -  \bi  \frac{\sqrt{6}}{3} 
    \frac{\alpha}{m_q} \expfac  \tau_{_N} 
    \trija{1}{M_X\!\!-\!m_{_N}}{\hf}{m_{_N}}{J_X}{M_X} 
   \vec{e}\,^*_{M_X-m_{_N}},\nonumber\\
&&\\
\label{SIGMA1}
\bSigma_{_{N(^28)}}(\bq)& = & \bi  \frac{\sqrt{2\pi}}{3} 
    \frac{q}{\alpha} \expfac 
   (1+4\tau_{_N}) 
   \sum_{\mu} 
    \trija{1}{M_X\!\!-\!\mu}{\hf}{\mu}{J_X}{M_X}\nonumber\\
&&\times Y^*_{1M_X-\mu}(\hat{q}) \cb{\mu},\\
\label{SIGMA2} 
\bSigma_{_{N(^48)}}(\bq) & = & -  \bi  \frac{\sqrt{\pi}}{6} 
    \frac{q}{\alpha} \expfac 
  (1-2\tau_{_N})  
   \sum_{\mu} 
 \trijb{1}{M_X\!\!-\!\mu}{\thf}{\mu}{J_X}{M_X}\nonumber\\ 
&&\times Y^*_{1M_X-\mu}(\hat{q})  \ca{\thf}{\mu},\\
 \bSigma_{_{\Delta}}(\bq) & = & \expfac 
   \ca{\thf}{m_{\Delta}},\\
\bigrho_{_{\Delta(^{2}10)}}(\bq) & = &
\frac{1}{2\tau_{_N}} \bigrho_{_{N(^28)}}(\bq),\\
\bP_{_{\Delta(^{2}10)}}(\bq)& = & \frac{1}{2\tau_{_N}} \bP_{_{N(^28)}}(\bq),\\
\label{SIGMA3}
\bSigma_{_{\Delta(^{2}10)}}(\bq)
& = & \frac{-1}{1+4\tau_{_N}} \bSigma_{_{N(^28)}}(\bq),
\end{eqnarray}
   where $\vec e_m$ is the spherical basis vector and $\alpha$ the oscillator 
parameter. 
The eigenstates of the total spin $S$ of the three quarks have been denoted by
  $\chi^{3/2}_{\mu}$ for $S=3/2,$ and 
$\chi^\lambda_{\sigma_{N}}$ for $S=1/2,$ 
with $\lambda$ indicating symmetry under interchange of the (12) 
quark pair.
We note that the definitions for the overlap integrals introduced in    
Eqs.\ (\ref{RHO})--(\ref{SIGMA}) differ by a factor 3 from the
corresponding expressions in Eqs.\ (9)-(11) of Ref.\ \cite{LTG}.
In addition, 
we  found a different result in the explicit 
calculation of the
integral entering into  
$ \bSigma_{_{N(^28)}}(\bq) $ (the integral in Eq.\ (14) of Ref.\ \cite{LTG} 
is smaller than our result 
by a factor $2\sqrt{2}$),
while we agree with the 
results for the integrals contributing to the remaining terms 
given in Eqs.\ (12) and (13) and Eqs.\ (15)-(19) of Ref.\ \cite{LTG}.

\section{Polarizabilities in the nonrelativistic $1/M$ expansion}
\label{app2}

In this appendix we collect the results for the polarizabilities obtained
from the multipole expansion of the Compton tensor 
in the $1/M$  nonrelativistic limit.

The leading contributions corresponding to the terms in Eq.\ 
(\ref{tensor_nrLEADING})
read 
($q=|\vec{q}\,|$)

\begin{eqnarray}
P^{(01,01)S} 
&=&{1 \over 18} {1 \over \alpha^2}\expfac  
       \sumX {a_X^2\over \dMx}  Z_d^{S,J_X}  
        \left( 1+(-1)^S\right), \\
P^{(01,12)1}
& =&  {1 \over 18} \sqrt{3 \over 5}  
   {1 \over m_q\alpha^2} \expfac  
     \sumX  a_X^2 {(-1)^{I_x-1/2} \over 2I_x}
       {  Z_{ad}^{2,S,J_X} \over \dMx },\\
P^{(11,11)S}_{para}&=&
{4 \over 27}
{1 \over m_q^2}\expfac
  { Z_\Delta^S\over M-M_\Delta }(1+(-1)^S),
\\
P^{(11,11)S}_{dia} &=& \delta_{S0}{1 \over 3\sqrt{6}} 
      {1 \over m_q\alpha^2}\expfac,  
\\
\hat{P}^{(01,1)S}_{\Sigma}&=&-{1 \over 36\sqrt{3}}
   {1 \over m_q\alpha^2} \expfac  
   \sumX a_X^2 {(-1)^{I_x-1/2} \over 2I_x}
    { Z_{ad}^{1,S,J_X}  \over \dMx }
(1-(-1)^S) .
\end{eqnarray}

Comparing these results with Eqs.\ (\ref{p0101_s})--(\ref{p011_s_f})
for the leading contributions of the polarizabilities in the convention of
Ref.~\cite{LTG}, we notice that $P^{(01,01)S}$,  $P^{(11,11)S}_{para}$,
and $P^{(01,12)1}$ involve the same matrix elements of the 
current in both expansion schemes,
while for the propagator in the crossed channel we now have 
$(M - M_X)^{-1}$ instead of $[E(\vec q\,)-E_X]^{-1}$.

The diamagnetic contribution to $P^{(11,11)0}$ from the two-photon 
interaction does not change in the $1/M$ expansion.

The polarizability $\hat{P}^{(01,1)S}$ has a leading contribution only 
from the spin-dependent term of the current,
$\hat{P}^{(01,1)S}_\Sigma,$ while the contribution from the 
convective current corresponds to $1/M$ corrections 
that are taken into account in the recoil term.

The recoil contributions can be obtained from the respective terms of 
Eq.\ (\ref{tensor_nrRECOIL}),

\begin{eqnarray}
\hat{P}^{(01,1)S}_{recoil}
& =&{\sqrt{2} \over 108\sqrt{3}}
    {1 \over m_q\alpha^2}\expfac \sumX a_X^2 
      {Z_d^{S,J_X} \over \dMx } (1+(-1)^S),\\
& &\nonumber\\
P^{(11,00)1}_{recoil} &=& -  {1 \over \sqrt{3}\, }
      {q^2\over  m_q } \,\expfac
       \sumX {a_X^2 Z_{1100}^{J_X} \over M [\dMx]^2 }  
       \left[1-  {M [\dMx] \over 3\alpha^2 }\right],
\\
& &\nonumber\\
P^{(11,02)1}_{recoil} &=&   - {1 \over \sqrt{3}\,} {1\over m_q }\,\expfac
       \sumX {a_X^2 Z_{1102}^{J_X}\over M [\dMx]^2 }  
       \left[1-  {M [\dMx] \over 3\alpha^2 }\right], 
\\
& &\nonumber\\
 P^{(11,11)S}_{recoil,C}& =& 
 {1\over 18\,}
 {\alpha^2 \over  m_q^2 } \,\expfac
       \sumX { a_X^2 Z_{1111,F}^{S,J_X} \over M [\dMx]^2 }
       \left[1-  {2 M [\dMx] \over 3\alpha^2 }\right], 
\\
& &\nonumber\\
 P^{(11,11)S}_{recoil,\Sigma} &=& {1\over 18\,}{
q^2 \over  m_q^2 } 
\,\expfac
       \sumX a_X^2 {(-1)^{I_x-1/2} \over 2I_x}
     {Z_{1111,\Sigma}^{S,J_X} \over M [\dMx]^2 }\nonumber\\
& &      \times \left[1-  {M [\dMx] \over 3\alpha^2 }\right], 
\\
& &\nonumber\\
\hat{P}^{(11,2)1}_{recoil,C} &=& 
 -{1 \over 6\sqrt{5}\,}{1 \over m_q^2 }\,\expfac
       \sumX {a_X^2  Z_{1102}^{J_X} \over M [M-M_X]^2 },   
\\
& &\nonumber\\
\hat{P}^{(11,2)1}_{recoil,\Sigma} &=&  {1 \over 2\sqrt{5}\,}
{1 \over m_q^2 }\,\expfac
       \sumX 
a_X^2 {(-1)^{I_x-1/2} \over 2I_x}
{Z_{1102}^{J_X} \over M [\dMx]^2 }\nonumber\\  
  & &   \times   \left[1-  {M [\dMx] \over 3\alpha^2 }\right]. 
\end{eqnarray} 

Comparing with the results in Eqs.\ (\ref{p1100_1})--(\ref{p112_1_sigma}), 
we find the same expressions for $P^{(11,00)1}$, $P^{(11,02)1}$, 
and $P^{(11,11)1}$ after the substitutions 
$[E(q)-E_X(q)]^{-1}\rightarrow [M-M_X]^{-1}$
and $E_X^{-1}(q) [E(q)-E_X(q)]^{-2}\rightarrow M^{-1}[M-M_X]^{-2}$.

As noticed previously, the additional recoil term for 
$\hat P^{(01,1)1}$ corresponds to the  $\hat P^{(01,1)1}_F$
term in Eq.\ (\ref{p011_s_f}).
Furthermore, $\hat P^{(11,2)1}_{recoil,C}$  
 does not contain the contribution 
from the convective current at the virtual-photon vertex resulting from
the term proportional to 
$J^\nu_{c,fX}(\vec q) \delta J^\mu_{c, Xi}(\vec q\, ')$ 
in Eq.\ (\ref{eq:recoil2}), because this contribution corresponds to a 
higher-order correction in $1/M$.

\newpage

\clearpage

\begin{table}[ht]
\label{table1}
\caption{Mixing parameters for the $^2\underline{8}$ and
$^2\underline{10}$ 
representation of the P-wave baryon resonances.}
\begin{center}
{\begin{tabular}{|c|c|c|c|c|c|c|}
$X$                                                        & 
$N(\frac{1}{2}^-,1535)$      & $N(\frac{1}{2}^-,1650)$     & 
$N(\frac{3}{2}^-,1520)$      & $N(\frac{3}{2}^-,1700)$     &  
$\Delta(\frac{1}{2}^-,1620)$ & $\Delta(\frac{3}{2}^-,1700)$\\
\hline
$a_X$  & $0.85$ & $0.53$ & $0.99$ & $0.11$ & $1.0$ & $1.0$\\
\end{tabular}
}\end{center}
\end{table}

\begin{table}[ht]
\label{table2}
\caption{Angular coefficients for the leading contribution to 
the GPs.}
\begin{center}
{
\begin{tabular}{|c|c|c|c|c|c|c|c|}
$L$ & $S$ & $J_X$  &  $Z_d^{S,J_X}$  &  $Z_c^{S,J_X}$  & $Z_{ad}^{L,S,J_X}$      
  & $Z_{ac}^{L,S,J_X}$ & $Z_{\Delta}^{S}$  
\\ \hline
1 & 0 & 1/2  & $\sqrt{2/3}$ & $\sqrt{2/3}$ & $-2/\sqrt{3}$ 
  & $2/\sqrt{3}$  & $-$ 
\\ \hline
1 & 0 & $3/2$ & $2\sqrt{2/3}$ & $2\sqrt{2/3}$ & $2/\sqrt{3}$ & $-2/\sqrt{3}$ 
  & $\sqrt{6}$
\\ \hline
1 & 1 & 1/2  & 2/3    & $-2/3$   & $-2\sqrt{2}/3$   &$-2\sqrt{2}/3$ 
  &  $-$
\\ \hline
1 & 1 & 3/2 & $-2/3$    & 2/3   & $-\sqrt{2}/3$   &$-\sqrt{2}/3$ 
  & $-1$ 
\\ \hline
2 & 1 & 3/2 & $-$    &   $-$    & $\sqrt{30}/3$  &$\sqrt{30}/3$  & $-$
\\ 
\end{tabular}
}
\end{center}
\end{table}

\begin{table}[hb]
\label{table3}
\caption{
Angular coefficients for the recoil contribution to the GPs.}
\begin{center}
{\begin{tabular}{|c|c|c|c|c|c|}
$S$ &
$J_X$ &  
$Z^{J_X}_{1100}$ & 
$Z^{J_X}_{1102}$ & 
$Z^{S,J_X}_{1111,C}$&
$Z^{S,J_X}_{1111,\Sigma}$
\\
\hline
 0&
  1/2 & $-$ 
 & $-$ 
 &
$\sqrt{{2\over3}}$&
 $1\sqrt{6}$ \\
\hline
$0$&
$3/2$ & $-$
 &  $-$
 &
$2\sqrt{{2\over3}}$&
 $-1\sqrt{6}$ \\
\hline
$1$&
$1/2$ & 
$1/27$&  
$1/27\sqrt{2}$&
 $-1/3$ &
 $-1/6$ \\
\hline
$1$&
$3/2$ & 
$-1/27$&
$-1/27\sqrt{2}$&
 $~1/3$ &
$~1/6$ \\
\end{tabular}
}\end{center}
\end{table}

\begin{table}[ht]
\label{table4}
\caption{Structure functions $P_{LL}$, $P_{TT}$, and $P_{LT}$
in $\mbox{GeV}^{-2}$.
The two entries for $P_{TT}$ and $P_{LT}$ originate from the first
and second equations of Eqs.\ (\ref{PTT}) and (\ref{PLT}), respectively. 
} 
\begin{center}
{\begin{tabular}{|c|c|cc|cc|}
&$P_{LL}$&\multicolumn{2}{c|}{$P_{TT}$}&\multicolumn{2}{c|}{$P_{LT}$}\\ \hline
$|\vec{q}\,|=0$ MeV
&37.0
&\multicolumn{2}{l|}{$-0.1\quad\vline\quad0.0$} 
&\multicolumn{2}{r|}{$-11.2\quad\vline\quad-15.8$}
\\ \hline
$|\vec{q}\,|=240$ MeV
&28.7
&\multicolumn{2}{r|}{$-2.8\quad\vline\quad-1.4$}
&\multicolumn{2}{r|}{$-8.8\quad\vline\quad-12.4$}
\\ \hline
$|\vec{q}\,|=600$ MeV
&9.9
&\multicolumn{2}{r|}{$-5.8\quad\vline\quad-3.1$}
&\multicolumn{2}{c|}{$-3.2\quad\vline\quad-4.5$}
\\
\end{tabular}
}\end{center}
\end{table}

\begin{table}[ht]
\label{table5}
\caption{Structure functions $P_{LL}-P_{TT}/\epsilon$ and $P_{LT}$
for $|\vec{q}\,|=600$ MeV in GeV$^{-2}$ ($\epsilon=0.62$).
The two entries for $P_{LL}-P_{TT}/\epsilon$ and $P_{LT}$ in our
calculation originate from the first and second equation of Eqs.\ (\ref{PTT})
and (\ref{PLT}), 
respectively. 
} 
\begin{center}
{\begin{tabular}{|c|cc|cc|}
&\multicolumn{2}{c|}{$P_{LL}-P_{TT}/\epsilon$}&
\multicolumn{2}{c|}{$P_{LT}$}\\ \hline
This calculation 
&\multicolumn{2}{c|}{$19.2\quad\vline\quad 14.9$}
&\multicolumn{2}{c|}{$-3.2\quad\vline\quad -4.5$}
\\ \hline
NRCQM of \cite{LTG}
&\multicolumn{2}{c|}{11.1}
&\multicolumn{2}{c|}{$-3.5$}\\ \hline
Experiment \cite{Roche_2000}
&\multicolumn{2}{c|}{$23.7$}
&\multicolumn{2}{c|}{$-5.0$}\\
&
\multicolumn{2}{c|}{$\pm2.2\pm0.6\pm 4.3$} 
&\multicolumn{2}{c|}{$\pm 0.8\pm 1.1 \pm 1.4$}\\
\end{tabular}
}\end{center}
\end{table}

\begin{figure}[ht]
\label{fig1}
\vspace{1cm}
\epsfxsize=13cm
\centerline{\epsffile{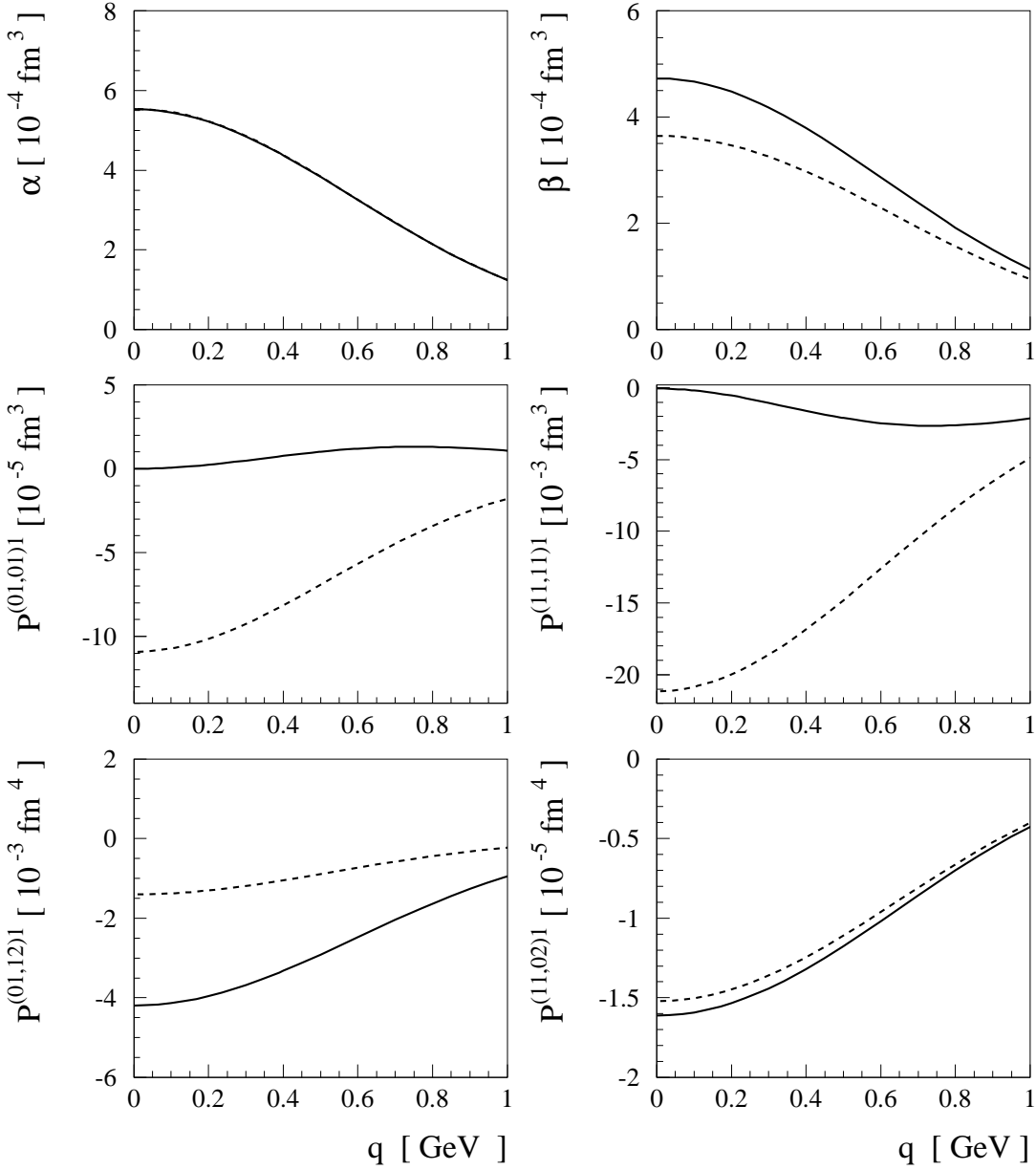}}
\caption[]
{GPs in the NRCQM as function of virtual-photon momentum 
$\mbox{q}=|\vec{q}\,|$.
Full lines: our results in the scheme of LTG, 
taking account of leading and recoil terms; 
dashed lines: calculation of Ref.~\cite{LTG}.
Note that the two calculations coincide in the case of $\alpha$.}
\end{figure}

\begin{figure}[ht]
\label{fig2}
\vspace{1cm}
\epsfxsize=13cm
\centerline{\epsffile{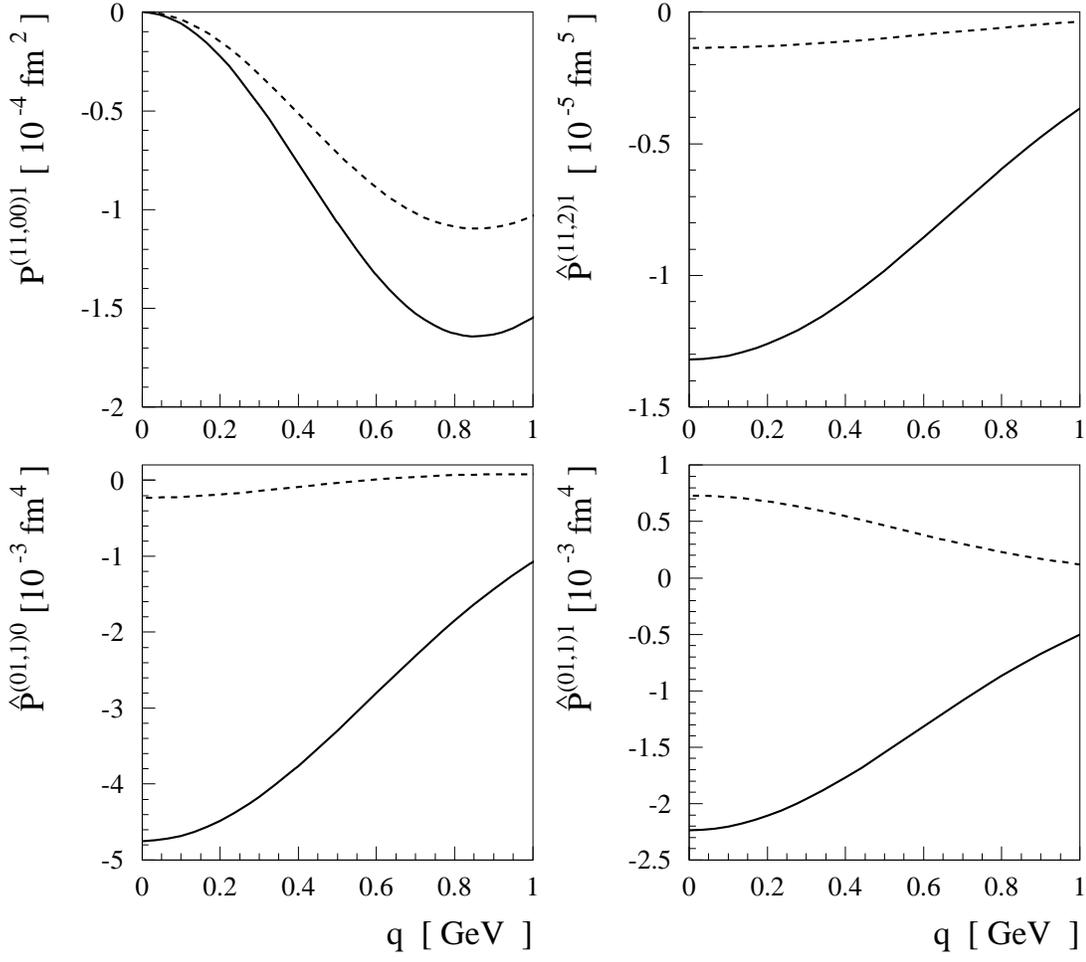}}
\caption[]
{GPs in the NRCQM as function of virtual-photon momentum 
$\mbox{q}=|\vec{q}\,|$.
Full lines: our results in the scheme of LTG, 
taking account of leading and recoil terms; 
dashed lines: calculation of Ref.~\cite{LTG}.}
\end{figure}

\begin{figure}[ht]
\label{fig3}
\vspace{1cm}
\epsfxsize=13cm
\centerline{\epsffile{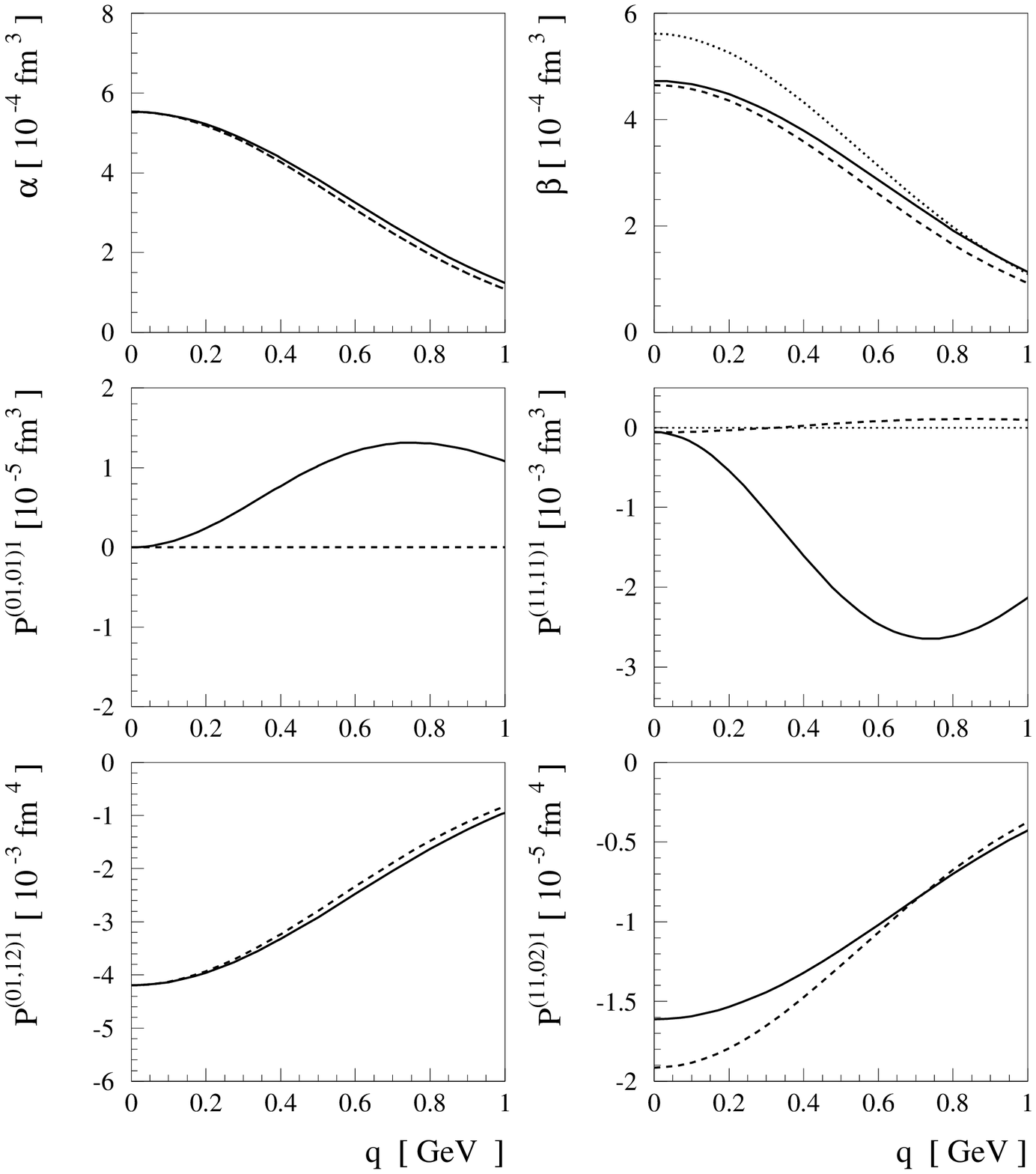}}
\caption[]
{GPs in the NRCQM as function of virtual-photon momentum 
$\mbox{q}=|\vec{q}\,|$.
Full lines: 
our results in the scheme of LTG, including the leading 
and recoil contributions;
dashed lines: nonrelativistic $1/M$ expansion of the
Compton tensor, obtained from the sum of the leading and recoil terms;
dotted lines: leading contributions.
   Note that dotted lines are only included if the leading contribution
is finite and different from the total result.
    To the order considered, the polarizabilities $\alpha$, 
$P^{(01,12)1}$, and $P^{(01,01)1}$ 
receive only contributions from the leading terms of Eq.\
(\ref{tensor_nrLEADING}), 
whereas $P^{(11,02)1}$ consists only of a recoil contribution 
from Eq.\ (\ref{tensor_nrRECOIL}).
}
\end{figure}

\begin{figure}[ht]
\label{fig4}
\vspace{1cm}
\epsfxsize=13cm
\centerline{\epsffile{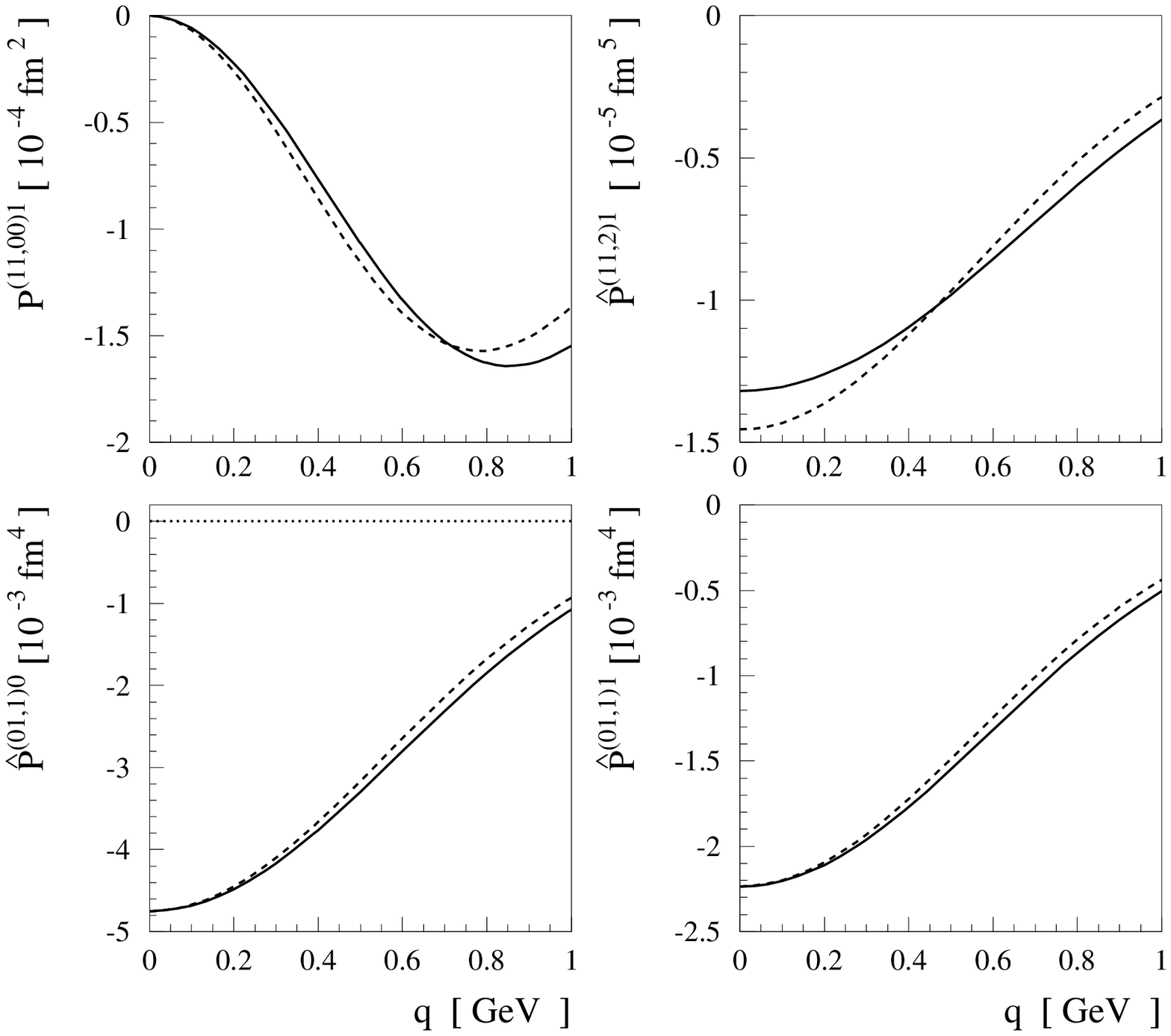}}
\caption[]
{GPs in the NRCQM as function of virtual-photon momentum 
$\mbox{q}=|\vec{q}\,|$.
Full lines: 
our results in the scheme of LTG, including the leading 
and recoil contribution;
dashed lines: nonrelativistic $1/M$ expansion of the
Compton tensor, obtained from the sum of leading and recoil terms;
dotted lines: leading contributions.
   Note that dotted lines are only included if the leading contribution
is finite and different from the total result.
    To the order considered, $\hat P^{(01,1)1}$ receives only a
contribution from the leading terms of Eq.\
(\ref{tensor_nrLEADING}), 
whereas $P^{(11,00)1}$ and  $\hat P^{(11,2)1}$
consist only of a recoil contribution 
from Eq.\ (\ref{tensor_nrRECOIL}).
}
\end{figure}

\begin{figure}[ht]
\label{fig5}
\vspace{1cm}
\epsfxsize=13cm
\centerline{\epsffile{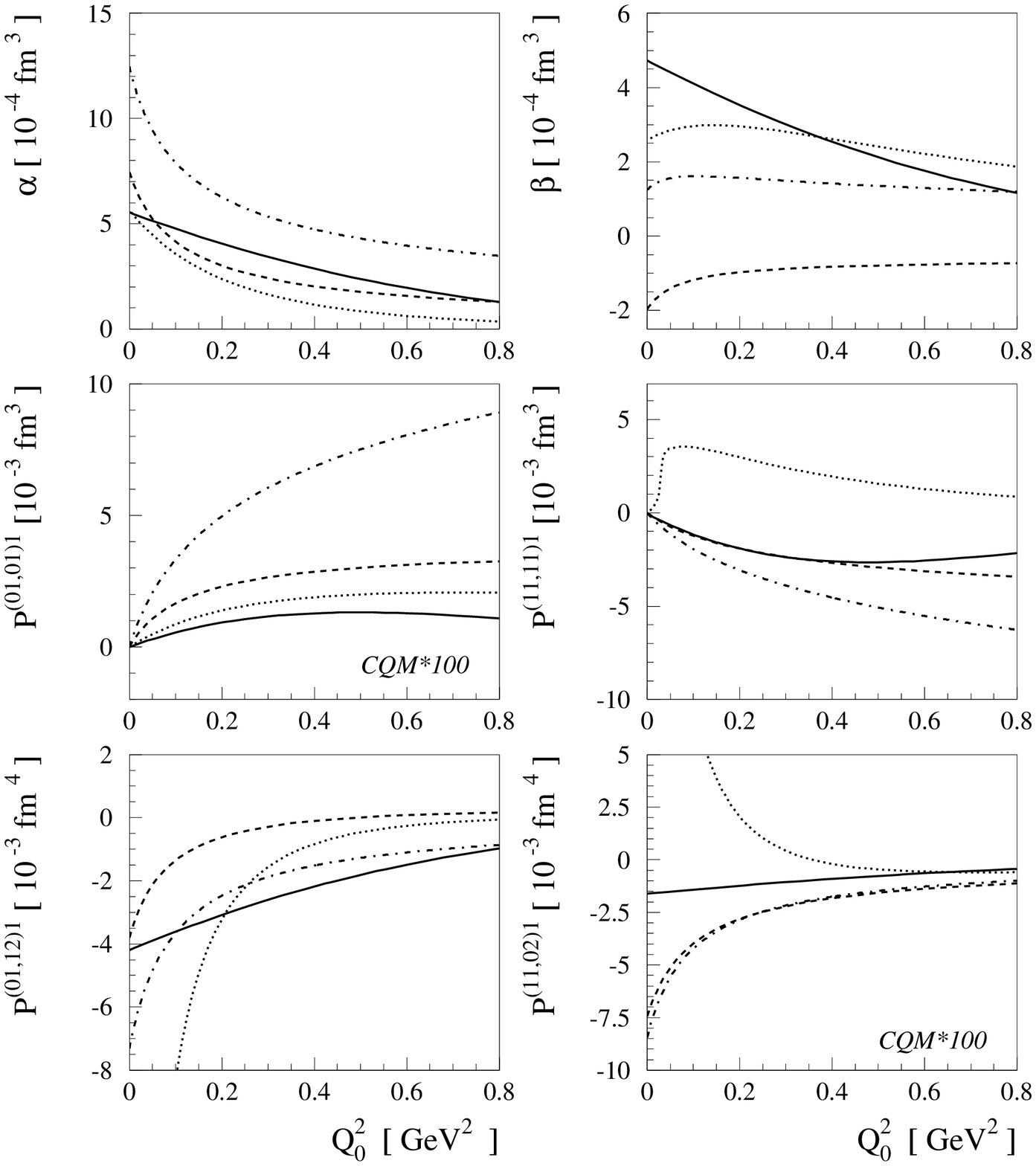}}
\caption[]
{Results for GPs in different model calculations as function of squared 
momentum transfer at $\omega'=0$, $Q^2_0=Q^2|_{\omega'=0}$.
Full lines: 
our results in the NRCQM with the scheme of LTG;
dashed lines: linear sigma model \cite{Metz_96};
dotted lines: effective Lagrangian model 
~\cite{Korchin_1998};
dashed-dotted lines :  heavy-baryon chiral perturbation theory
\cite{Hemmert_2000}.}
\end{figure}

\begin{figure}[ht]
\label{fig6}
\vspace{1cm}
\epsfxsize=13cm
\centerline{\epsffile{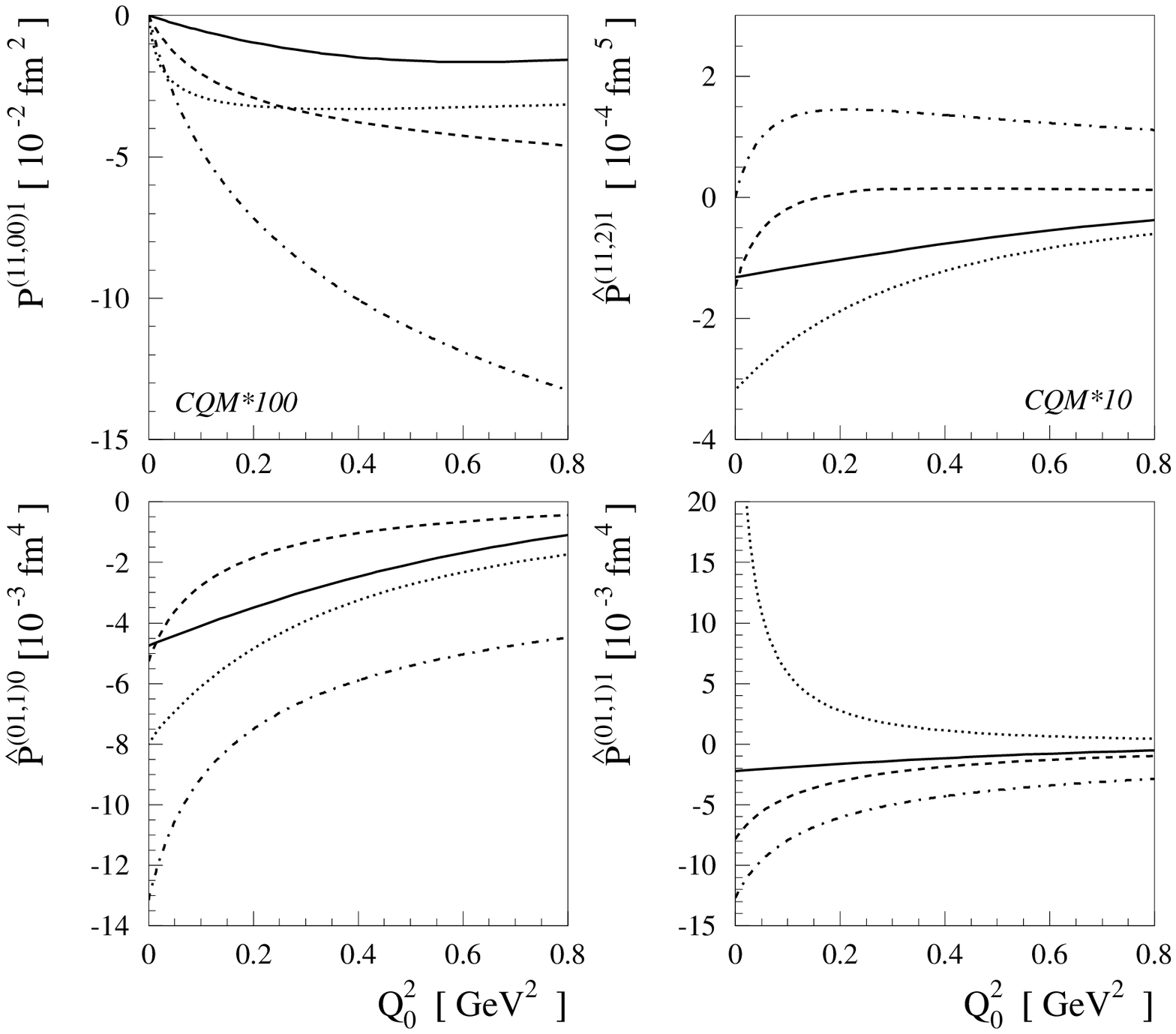}}
\caption[]
{Results for GPs in different model calculations
as function of squared momentum transfer at 
$\omega'=0$, $Q^2_0=Q^2|_{\omega'=0}$.
Full lines: our results in the NRCQM with the scheme of LTG;
dashed lines: linear sigma model \cite{Metz_96};
dotted lines: effective Lagrangian model \cite{Korchin_1998};
dashed-dotted lines : heavy-baryon chiral perturbation theory
\cite{Hemmert_2000}.
}
\end{figure}

\begin{figure}[ht]
\label{fig7}
\vspace{1cm}
\epsfxsize=13cm
\centerline{\epsffile{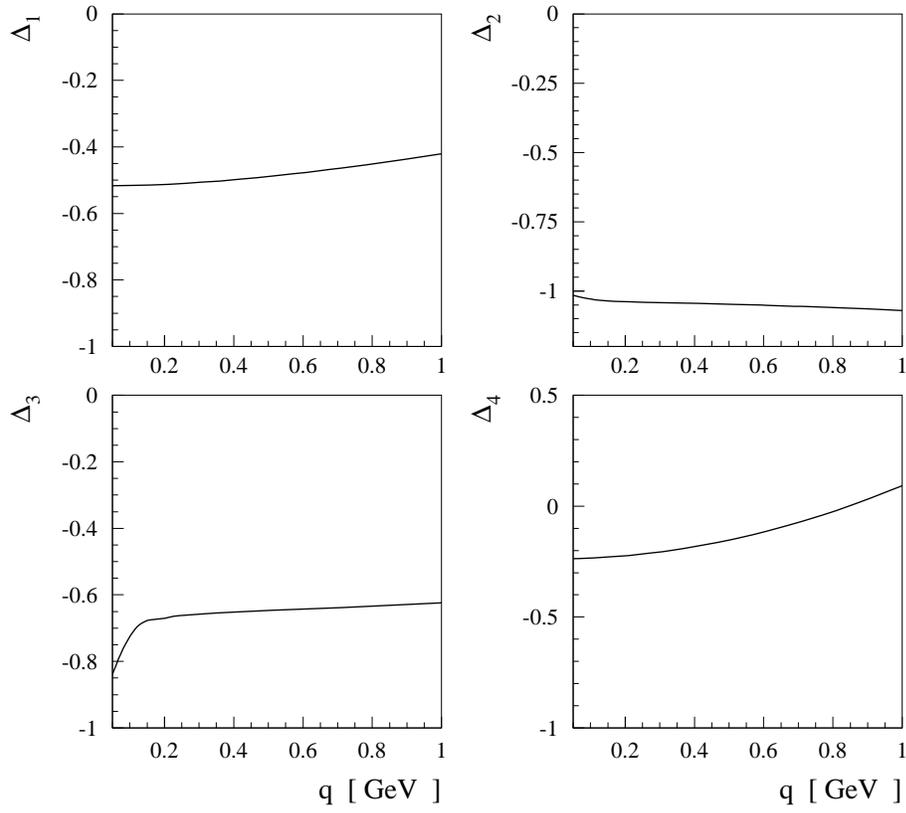}}
\caption[]
{Violation of the relations of Eqs.\ (\ref{eq:rel1}) - (\ref{eq:rel4})
as function of $\mbox{q}=|\vec{q}\,|$.
}
\end{figure}
\end{document}